\documentstyle{article}
\def\ut#1{\mathop{\vtop{\ialign{##\crcr
     $\hfil\displaystyle{#1}\hfil$\crcr\noalign
     {\kern1pt\nointerlineskip}\hbox{$\hfil\sim\hfil$}\crcr
     \noalign{\kern1pt}}}}}

\def\undersymbol#1#2{\mathop{\vtop{\ialign{##\crcr
     $\hfil\displaystyle{#2}\hfil$\crcr\noalign
     {\kern1pt\nointerlineskip}\hbox{$\hfil#1\hfil$}\crcr
     \noalign{\kern1pt}}}}}

\begin{document}
\begin{titlepage}
\begin{flushright}
Lecce University LE/ASTRO 3/96\\
Pavia University FNT/T-96/16\\
Zurich University ZU-TH 8/96\\
\end{flushright}
\vfill
\begin{center}
{\large\bf Halo Dark Clusters of Brown Dwarfs and Molecular Clouds}
\vskip 0.5cm
F.~De Paolis$^{1,2}$,
G.~Ingrosso$^{3}$,
Ph.~Jetzer$^{2}$
and M.~Roncadelli$^{4,*}$
\vskip 0.5cm
$^1$
Bartol Research Institute, University of Delaware, Newark, Delaware,
19716-4793, USA\\
$^2$
Paul Scherrer Institut, Laboratory for Astrophysics, CH-5232 Villigen PSI,
and
Institute of Theoretical Physics, University of Zurich, Winterthurerstrasse
190, CH-8057 Zurich, Switzerland\\
$^3$
Dipartimento di Fisica, Universit\`a di Lecce, Via Arnesano, CP 193, 73100
Lecce, Italy\\
and
INFN, Sezione di Lecce, Via Arnesano, CP 193, 73100 Lecce, Italy\\
$^4$
INFN, Sezione di Pavia, Via Bassi 6, I-27100, Pavia, Italy
\end{center}
\vfill 
 
\begin{abstract}
The discovery of Massive Astrophysical Compact Halo Objects (MACHOs)
in microlensing experiments makes it compelling to understand
their physical nature, as well as their formation mechanism.
Within the present uncertainties, 
brown dwarfs are a viable candidate for MACHOs, and the present paper
deals with this option.
According to a recently proposed scenario, brown dwarfs are clumped
along with cold molecular clouds into 
dark clusters -- in several respects similar 
to globular clusters -- which 
form in the outer part of the galactic halo.
Here, we analyze the dynamics of these dark clusters  and we address the
possibility that a sizable fraction of MACHOs can be binary
brown dwarfs. 
%Remarkably enough, it turns out that 
%the presence of cold molecular clouds can be tested experimentally,
%for they should give rise to absorption lines in the spectra of 
%stars which lie close to a previously microlensed one.
Moreover, we point out that Ly-$\alpha$
absorption systems naturally fit within the present picture.\\

\noindent{Subject headings:
Dark matter - Galaxy: halo - Galaxy: kinematics and dynamics - 
Gravitational microlensing}
\end{abstract}
\vfill

$^*$ Work partially supported by Dipartimento di Fisica Nucleare
e Teorica, Universit\`a di Pavia, Pavia Italy
\end{titlepage}
\newpage 
\baselineskip=21pt

\section{Introduction}
Observations of microlensing events (Alcock et al. \cite{alcock}, Aubourg et
al. \cite{aubourg}) towards the Large Magellanic Cloud (LMC)
strongly suggest that a substantial fraction of the galactic halo
should be in the form of dark compact objects, named MACHOs (massive 
astrophysical compact halo objects)
(De R\'ujula, Jetzer and Mass\`o \cite{der}).
 
Actually, the MACHO collaboration has recently
announced the discovery of several new events during
their second year of observations (Alcock et al. \cite{Alcock1}), and
8 microlensing events have been detected so far \footnote{It should
be mentioned that the MACHO team has found at least
seven more events (which are reported on the Alert list), but
a full analysis of them has not yet been published.}.
 
Although the presently-available limited statistics prevents clear-cut
conclusions to be drawn from experimental data
(Gates, Gyuk and Turner \cite{gates}), the evidence for such a
discovery is firm and its implications are striking. In fact  -- under
the assumption that MACHOs are indeed located in the galactic halo --
the inferred halo mass in MACHOs within 50 kpc turns out to be 
$2.0^{+1.2}_{-0.7}\times 10^{11}~M_{\odot}$ (Alcock et al. [1997]),
which is several times larger than the mass of all known stellar
components of the Galaxy and represents a relevant portion of the
galactic dark matter. Remarkably enough, this result is almost
independent of the assumed galactic model. Unfortunately, this
circumstance contrasts with the strong model-dependence of the average
MACHO mass. 
It has become customary
to take the standard spherical halo model as  a baseline for comparison.
Regretfully -- because of the low statistics --
different data-analysis procedures
lead to results which are only marginally consistent. 
Specifically, within the standard halo model, the average MACHO mass 
reported by the MACHO team is
$0.46^{+0.3}_{-0.2}~M_{\odot}$ (Alcock et al. [1997]), whereas the mass 
moment method (De R\'ujula, Jetzer and Mass\'o [1991]) yields 0.27 $M_{\odot}$
(Jetzer [1996]).

What can be reliably concluded from the existing data-set is that MACHOs
should lie in the mass range $0.05 - 1.0 ~M_{\odot}$
(see also Table 9 in Alcock et al. [1997]), but stronger claims 
are unwarranted because of the high-sensitivity of the average 
MACHO mass on the uncertain 
properties of the considered galactic model (Evans [1996], De Paolis,
Ingrosso and Jetzer [1996]). 

Mass values $> 0.1~M_{\odot}$  suggest
that MACHOs should be either M-dwarfs or else white dwarfs. 
Observe that these mass values naturally arise within the standard
halo model.

As a matter of fact, deeper considerations show that the M-dwarf option
can look problematic. The null results of several searches for low-mass 
stars both in the disk and in the halo of our galaxy (Hu et al. \cite{hu})
entail that the halo cannot be mostly in the form of hydrogen-burning
main-sequence M-dwarfs.
Optical imaging of high-latitude fields taken with
the Wide Field Camera of the Hubble Space Telescope
indicates that less than $\sim 6~\%$ of the
halo mass can be in this form (Bahcall et al. \cite{bahcall}). 
A more detailed analysis which accounts for the fact that halo
stars likely have a lower metallicity (with respect to the solar
one) leads to an even more stringent upper limit of less
than $\sim 1~\%$ (Graff and Freese \cite{graff}).
We emphasize that these results are derived under the assumption of a smooth spatial
distribution of M-dwarfs, and become less severe in the case of a
clumpy distribution (Kerins \cite{kerins}).
In the latter case, as pointed out by Kerins \cite{kerins97},  
the dynamical
limits and HST observations require that
the overwhelming fraction of M-dwarfs, at least 95\%, 
must still reside in clusters at present. 
His analysis shows that there exists a wide 
range of cluster masses and radii which 
are consistent with these requirements.

As we said, an alternative explanation for 
MACHOs can be provided -- within the standard
spherical halo model -- 
by white dwarfs, and a scenario with white dwarfs as a
major constituent of the galactic halo dark matter has been explored
(Tamanaha et al. \cite{tama},
Fields et al. \cite{fields}, Adams and Laughlin \cite{adams},
Chabrier, Segretain and M\'era \cite{chabrier}).
However, even this proposal meets difficulties.
Besides requiring a rather {\it ad hoc} initial mass function
(IMF) of the progenitor stars
sharply peaked somewhere in the range $1 - 8~ M_{\odot}$
and a halo age larger than $\sim 16$ Gyr, 
strong constraints on the number density of halo white dwarfs arise from
present-day metal abundances in the interstellar medium
(Ryu, Olive and Silk \cite{ryu},
Gibson and Mould [1996]) and from deep galaxy counts
(Charlot and Silk \cite{charly}).
In any case, future HST deep field exposures will either find the white dwarfs or 
put constraints on their fraction in the halo (Kawaler \cite{kawaler}).

Mass values $\ut < 0.1~M_{\odot}$ make brown dwarfs 
\footnote{Brown dwarfs have been discovered quite recently 
in the solar neighbourhood and in the Pleiades cluster (Rebolo et al.
\cite{rebolo}, Nakajima et al. \cite{nakajima}).
The idea that MACHOs are brown dwarfs has been contemplated by several
authors (see Carr \cite{bcarr} and references therein).}
an attractive candidate for MACHOs
\footnote{We notice that the limit for hydrogen burning -- usually quoted as
$0.08~M_{\odot}$ -- gets increased up to $0.11~M_{\odot}$ for
low-metallicity objects, such as a halo population (D'Antona [1987], 
Burrows, Hubbard \& Lunine [1989]).}. 
In fact, these mass values are supported by several 
nonstandard halo models. An example is provided by
the maximal disk model
(van Albada \& Sancisi [1986]; Persic \& Salucci [1990]; Sackett
[1996]), 
in which more matter is comprised within the disk whereas the halo
is less massive as compared with the standard halo model.
The latter fact implies a falling rotation curve, and so a smaller
transverse velocity of MACHOs. Hence, the microlensing time scale gets
longer for a given MACHO mass, which means a smaller implied MACHO
mass for a given observed timescale. We stress that a reduced
transverse velocity of MACHOs arises also in other nonstandard halo
models. For instance, a radially anisotropic velocity distribution or
a halo rotation would do the job
(Alcock et al. [1997]; Evans [1996]; 
De Paolis, Ingrosso \& Jetzer [1996]).
We also notice that the EROS collaboration (Renault et
al. \cite{renault}) has shown that MACHOs in the mass range
$10^{-7} - 2 \times 10^{-2}~M_{\odot}$ do not
contribute significantly (less than 20\%) to the halo dark matter
(this result is consistent with the MACHO experiment for objects of
mass $0.1 M_{\odot}$ to $1 M_{\odot}$).

Although present uncertainties do not permit to make any sharp statement
about the nature of MACHOs, brown dwarfs still look as a viable
possibility to date, and we shall stick to it throughout.

Even if MACHOs are indeed brown dwarfs,
the problem nevertheless remains to explain their formation,  
as well as  the nature of the remaining dark matter in galactic 
halos.

We have previously proposed a scenario in which dark clusters 
of brown dwarfs 
and cold molecular clouds -- mainly of $H_2$ -- naturally form in 
the halo at galactocentric distances larger than $10-20$ kpc (De Paolis et 
al. \cite{depaolis1}-\cite{depaolis4}). 
Similar ideas have also been put forward by Gerhard and Silk \cite{Gerhard}.
A slightly different picture based on the presence of a strong cooling-flow
phase during the formation of our galaxy has been considered by Fabian
and Nulsen \cite{fn,nf} and leads to a halo made of low-mass objects.
In addition, Pfenniger, Combes and Martinet \cite{pfenniger}
suggested that $H_2$ clouds can constitute the dark matter in the disk
of our galaxy.

The model in question encompasses the one first proposed by Fall and Rees 
\cite{Fall} to explain the formation of globular clusters,
and no substantial additional hypothesis is 
required.
Various resulting observational implications 
have also been addressed. In particular\\
$i$) the $\gamma$-ray flux arising from halo molecular clouds through the 
interaction with high-energy cosmic-ray protons has been estimated 
(De Paolis et al. \cite{depaolis1,depaolis2});\\
$ii$) an anisotropy in the Cosmic Background Radiation (CBR) is predicted 
to show up when looking at the halo of M31 galaxy (De Paolis et al. 
\cite{depaolis3});\\
$iii$) the infrared emission from MACHOs located in the halo of M31 galaxy 
turns out to be observable with the detector on
ISO satellite or with the 
next generation of 
satellite-borne detectors (De Paolis et al. \cite{depaolis3}).

We would like to stress  that a large amount 
of MACHOs -- up to $50\%$ in mass -- can well
consist of binary brown dwarfs, formed either by the same fragmentation process
that produces individual brown dwarfs or
later when the dark clusters start to undergo core collapse.

The aim of the present paper is to discuss in a systematic
fashion further aspects of the above scenario.
Basically, we try to 
figure out the dynamics of dark clusters. More specifically,
we investigate the constraints which ensure their survival
against various kinds of gravitational perturbations.
In addition, we demonstrate that -- because of dynamical friction on
molecular clouds in the dark cluster cores (to be referred to as
{\it frictional hardening}) -- the present orbital radius of not too hard
binary brown dwarfs turns out to be typically of the order of 
the Einstein radius for microlensing towards the LMC.
As a consequence, taking also into account the adopted 
selection procedure in the data analysis, 
we understand why they have not been resolved so far -- 
still, we argue that they can be resolved in future microlensing
experiments with a more accurate photometric observation.
Finally, we show that Ly-$\alpha$ absorption
systems naturally fit within our 
model.

The plan of the paper is as follows. In Section 2 we recall
the main points of the considered picture for the formation of
dark clusters. Various dynamical constraints 
are thoroughly analyzed in Section 3, paying 
particular attention to the phenomenon of core collapse. 
In Section 4 we study the process whereby brown dwarfs form 
close binary systems (as a consequence of core collapse) and we investigate
the mechanism of frictional hardening in great detail.
In Section 5 we turn our attention to the thermal balance
in the halo molecular clouds. 
Section 6 contains a short discussion
of the relevance of Ly-$\alpha$ absorption systems for the present scenario.
Our conclusions are offered in Section 7.

\section{Scenario for dark cluster formation}
As shown elsewhere (De Paolis et al. \cite{depaolis1,depaolis2}), 
the model in question encompasses
the one first considered by Fall and Rees 
\cite{Fall} for the formation of globular clusters 
\footnote{A somewhat different extension of the Fall and Rees \cite{Fall}
scheme has been proposed by Ashman \cite{ashman}. Among his
motivations was the large spread in the age of globular clusters
found by Larson \cite{larson}. However, it seems nowadays that the
stars in the halo of our galaxy have a small scatter in ages
(Unavane, Wyse and Gilmore \cite{uwg}).}  and relies 
upon the result of Palla, Salpeter and Stahler \cite{Palla} (hereafter PSS)
that the lower bound on the Jeans mass in a collapsing
metal poor cloud can be as low as $10^{-2} M_{\odot}$, provided
certain environmental conditions are met.
 
Let us begin by summarizing the ideas of Fall and Rees \cite{Fall}
from a point of view which is most convenient for our considerations.

After the initial collapse, the proto galaxy (PG) is expected to 
reach a quasi-hydrostatic equilibrium state with virial 
temperature $\sim 10^6$ K. Fall and Rees \cite{Fall} have shown
that in such a situation a thermal instability develops: 
density enhancements rapidly grow as the gas cools to lower temperatures.
In fact, irregularities in the inflow during the gas collapse and also
fluctuations in the distribution of nonbaryonic dark matter (if 
present on the galactic scale) would introduce perturbations with a wide
range in size and amplitude. As a result, randomly distributed
overdense regions will form inside the PG. For reasons
that will become clear later, these overdense regions will be referred
to as proto globular cluster (PGC) clouds.

Under the assumption that the plasma in the PG is in 
collisional ionization equilibrium,  it turns out that 
the cooling rate (as a function of density $\rho$ and temperature $T$)
has the form 
\begin{equation}
\Lambda(\rho, T) = \rho^2 L(T)~ \label{eqno:R1} 
\end{equation}  
(the expression of $L(T)$ can be found e.g. in Einaudi and Ferrara
\cite{Einaudi}). 
The cooling time is
\begin{equation}
t_{cool} = \frac{3 \rho k_B T}{2 \mu (\Lambda - \Gamma)}~,
\label{eqno:R2}
\end{equation}
where the heating rate $\Gamma$ due to external heating sources
has been taken into account
(here $\mu \simeq 1.22~ m_p$ is the mean molecular mass of the primordial
gas). Since at the high temperatures under consideration
the heating rate can safely be neglected, it follows 
that $t_{cool} \sim \rho^{-1}$.
On the other hand, the free-fall time reads
\begin{equation}
t_{ff} \sim (G \rho)^{-1/2}~, \label{eqno:R3}
\end{equation}
and so we see that $t_{cool}$ decreases faster than $t_{ff}$ as $\rho$
increases. As the above quasi-hydrostatic equilibrium state of the PG
is characterized by the condition $t_{cool} \sim t_{ff}$, it is clear
that inside the PGC clouds we have $t_{cool} < t_{ff}$. That is to say, the 
PGC clouds cool more rapidly than the rest of the PG. This process continues
until hydrogen recombination occurs,
because as soon as this happens 
-- at a temperature $\sim 10^4$ K --  
the cooling rate decreases precipitously, under the ionization
equilibrium assumption
(Dalgarno and  McCray \cite{Dalgarno}). 
Therefore, the regime $t_{cool} > t_{ff}$ should now be
established even in the PGC clouds and so the PG can be
regarded at this stage as a two-phase medium, with cold PGC clouds in pressure
equilibrium with the external (inter PGC clouds)
diffuse hot gas.

However, Kang et al. \cite{Kang}
realized that the fast radiative cooling of the PGC clouds (from 
$10^6$ K to $10^4$ K) implies that the plasma inside these clouds cools
more rapidly than it recombines, so that the above ionization
equilibrium assumption is violated. Actually, the 
out-of-equilibrium recombination results in an enhanced ionization
fraction. This fact does not affect the previous conclusions for
temperatures $> 10^4$ K, but entails drastic changes at lower
temperatures. Indeed, the existence of a sizable amount of 
protons and electrons at temperatures $< 10^4$ K gives rise to
$H_2$ formation via the following reactions
\footnote{We wish to emphasize that the familiar reaction 
$H+H \rightarrow H_2 + \gamma$ (which requires
grains in order to be efficient) is presently irrelevant, since no dust exists
in the metal poor PGC clouds.}
\begin{equation}
H+p \rightarrow H_2^+ + \gamma~,
~~~~~~~H+e \rightarrow H^- +\gamma
\label{eqno:R6}
\end{equation}
and
\begin{equation}
H^+_2 + H \rightarrow H_2 + p~,
~~~~~~~H+ H^- \rightarrow H_2 + e ~.
\label{eqno:R7}
\end{equation}
As a consequence, the PGC clouds undergo a further cooling below
$10^4$ K. Specifically, there is a direct radiative cooling
via reactions (\ref{eqno:R6}) and a radiative cooling via
excitation of roto-vibrational transitions  of $H_2$ (observe that $H_2$ 
is further
produced by reactions (\ref{eqno:R7}) and (\ref{eqno:R8}) below). 
We stress that the latter process 
is very effective (much more than the former) at temperatures
$< 10^4$ K and plays a crucial r\^ole in our considerations. 
Since now $t_{cool} \ll t_{ff}$ in the PGC clouds, the collapse
goes on and the PGC cloud  density
rises steadily. When the number density in the PGC
clouds exceeds $10^8~ {\rm cm^{-3}}$, the $H_2$ production
increases dramatically thanks to the three-body reactions
\begin{equation}
H+H+H \rightarrow H_2 + H~,
~~~~~~~~H+H+H_2 \rightarrow H_2 +H_2 ~,
\label{eqno:R8}
\end{equation}
as pointed out by PSS.
In effect, these reactions are so efficient that virtually all 
the atomic hydrogen gets rapidly converted 
\footnote{Observe that -- at variance with reactions (\ref{eqno:R6})
and (\ref{eqno:R7}) -- reactions (\ref{eqno:R8}) do not require the 
presence of electrons and protons as a catalyst.} to $H_2$. 
Correspondingly, the cooling of the PGC clouds is strongly enhanced
and their evolution can proceed as in the scenario
proposed by PSS.

Still, it goes without saying that $H_2$ can be dissociated by various sources
of ultraviolet (UV) radiation, like an active galactic nucleus
(AGN) or a population of massive young stars (population III, see Carr, Bond
and Arnett \cite{RCarr}) at the 
centre of the PG. So, the ultimate fate of the PGC clouds 
strongly depends -- apart from other environmental conditions (more
about this, later) -- on the survival of $H_2$.

In the early phase of the PG an AGN is expected to form 
at its centre  along with population III stars, 
because of the disruption of central PGC clouds. 
This indeed happens, for the cloud collision time is shorter than the
cooling time in the central region of the PG.

Thus, $H_2$ will be dissociated at galactocentric distances smaller than
a certain critical value $R_{crit}$. Following the analysis 
of Kang et al. \cite{Kang},
it is straightforward to evaluate $R_{crit}$. Consider first the case of 
a UV flux due to a central AGN. Then we find
\begin{equation}
R_{crit}^{AGN} \simeq 
\left(\frac{L_{AGN}}{2 \times 10^{42}~{\rm erg~ s^{-1}}
 }\right)^{1/2}~{\rm kpc}~. 
\label{eqno:3a}
\end{equation}
For typical luminosities up to $L_{AGN} \simeq 10^{45}$ erg s$^{-1}$,
eq. (\ref{eqno:3a}) yields
$R_{crit}^{AGN} \simeq 20$ kpc.
On the other hand, when the UV dissociating flux is produced by 
massive young stars mainly located at the centre of the PG,
the  critical galactocentric distance turns out to be 
\begin{equation}
R_{crit}^{\star}\simeq \left( \frac{10^{-3}~{\rm kpc^{-3}}}{n_0} \right) 
\left(\frac{L_{tot}}{L_{\star}} \right)
~{\rm kpc}~. \label{eqno:3b}
\end{equation}
In eq. (\ref{eqno:3b}) $L_{\star} \simeq 2\times 10^{38}$ erg s$^{-1}$ 
is the bolometric luminosity of a single B0 V star.
Assuming a total stellar luminosity 
$L_{tot}$ up to $\simeq 2 \times 10^{45}$ erg s$^{-1}$ and a central number
density  $n_0$ 
up to  $\simeq 10^3~ {\rm kpc}^{-3}$, we find $R_{crit}^{\star} 
\simeq~ 10$ kpc.
In conclusion, $H_2$ should remain undissociated at galactocentric 
distances larger than $10 - 20$ kpc.

\subsection{Globular clusters}
According to the foregoing analysis, in the inner galactic halo
-- that is for galactocentric distances smaller than $10 - 20$ kpc --
$H_2$ gets dissociated \footnote{This indeed occurs for a wide
range of UV fluxes and PGC cloud densities (Kang et al. \cite{Kang}).},
thus preventing any further cooling of the PGC clouds below
$T \sim 10^4$ K. Therefore, these clouds remain for a long time in
quasi-hydrostatic equilibrium at $T \sim 10^4$ K (namely we
have $t_{cool} > t_{ff}$ during this period). In such a situation a 
characteristic mass scale gets imprinted on the PGC clouds
by the gravitational instability, thereby resulting in a strongly peaked 
mass spectrum of the PGC clouds. In fact, Fall and Rees \cite{Fall}
have shown that the Jeans mass of the PGC clouds after the long
permanence at $T \sim 10^4$ K is given (as a function of the
galactocentric distance $R$) by \footnote{We stress that the 
$R$-dependence in eqs. (\ref{eqno:M1}) and (\ref{eqno:M2}) holds
for $R <20$ kpc only (see the discussion in Vietri and Pesce 
\cite{Vietri}), while for $R> 20$ kpc the mass of globular 
clusters tends rather to decrease.}
\begin{equation}
M_{PGC}(R) \simeq 5\times 10^5~ (R/{\rm kpc})^{1/2} ~M_{\odot}~,
\label{eqno:M1}
\end{equation}
while the PGC radius turns out to be
\begin{equation}
r_{PGC}(R) \simeq 20~ (R/{\rm kpc})^{1/2}~{\rm pc}~.
\label{eqno:M2}
\end{equation} 

Observe that in the present case the propagation of sound waves erases
all large-scale perturbations, leaving only those at small scales.

Surely, this is not the end of the story, since the UV flux is expected to 
eventually decrease.
So, after some time a nontrivial fraction
of $H_2$ should anyway form, causing a sudden drop of the PGC cloud temperature
well below $\sim 10^4$ K
(the clouds now enter the regime 
$t_{cool} < t_{ff}$).
What next happens is a rapid growth of 
the small-scale perturbations, which lead
directly (in one step) -- due to the thermal
instability -- to formation of stars
inside the PGC clouds (Murray and Lin \cite{Murray}).
It goes without saying that we expect supernova explosions to have 
occurred, thus giving rise to shocks inside globular clusters  
and, therefore, leading to a rather wide stellar IMF.    
In this way, globular clusters should form as they are
observed today especially in the inner part of the galactic halo. 
Incidentally, the formation of the PGC clouds could have been 
delayed until after the Galaxy was
enriched by metals due to population III stars, in this manner explaining 
the absence of globular clusters with primordial metal 
abundances and the radial gradient of metallicity in the galactic halo.

\subsection{Dark clusters of MACHOs and cold molecular clouds}
As we have seen,
in the outer galactic halo -- namely for galactocentric distances 
larger than $10-20$ kpc -- $H_2$ molecules are not dissociated 
due to the absence of a significant UV flux.
Under the assumption of a quiet environment, this
circumstance entails a very efficient cooling of the PGC clouds, 
whose state is therefore characterized by the condition $t_{cool} < t_{ff}$.
Hence, the gravitational collapse is expected to occur as 
in the scenario of PSS. 
Specifically, both the temperature decrease and the density increase
cause a substantial drop of the Jeans mass for the PGC clouds.
As a consequence, they fragment into smaller and smaller clouds.
This process stops when the clouds become optically thick to their own
radiation, since then cooling manifestly gets uneffective.
PSS have shown that such a 
situation occurs when the Jeans mass is as low as $10^{-2}~M_{\odot}$.
We remark that the presence of rotation and magnetic fields in the 
PGC clouds would allow the Jeans mass to drop even further.
Obviously, the fragmentation down to low-mass values is favoured if 
the initial gas has been  metal enriched by population III 
stars (see Figure 7 in Palla and Stahler \cite{Palla1}).

The result of the above picture is the formation of dark clusters 
containing compact objects with an IMF 
peaked in the range $10^{-2}-10^{-1}~M_{\odot}$. This value
is close to the peak mass $\sim 0.1 ~M_{\odot}$ of the present
day IMF in the disk (Miller and Scalo \cite{miller}), 
but in our case we expect it
to decrease much more rapidly for higher masses and thus to be more
narrow around its mean value. 
So, the compact objects in question should be predominantly brown
dwarfs, even though a fraction of M-dwarfs can be present. Actually,
we want to stress that our considerations still hold true even if a
large fraction of MACHOs consists of M-dwarfs \footnote{The reader
should keep this point in mind, in spite of our taking about brown
dwarfs. Of course, quantitative changes in some results will occur,
because M-dwarfs are more massive than brown dwarfs (this is
especially true concerning Sects. 3 and 4).} (as already mentioned,
Kerins \cite{kerins97} has shown that M-dwarfs clumped into 
clusters is a viable possibility to date).
Notice that the expected lower limit of $\sim 10^{-2}~M_{\odot}$
is consistent with present bounds coming
from microlensing data, as found by the EROS collaboration
(Renault et al. \cite{renault}). 
 
What are the environmental conditions (besides $H_2$ survival) which
permit the mechanism in question to work? First, turbulence should be 
negligible. Indeed, turbulence effects would make fragments collide
and -- as shown by PSS -- 
this fact would increase the Jeans mass. Second, no sizable gravitational 
perturbations (shocks, supernova explosions, etc.) should be present
\footnote{This circumstance is consistent with our assumption that the
stellar IMF in the dark clusters is more sharply peaked than the IMF
in the globular clusters and in the disk.},
for otherwise they would induce the gravitational collapse before
cooling has succeeded in lowering the fragment Jeans mass down to
the above-mentioned values. 
Because these environmental conditions are likely to have 
occurred in the
outer halo, the mass of the produced compact objects 
is expected to be  close to the corresponding Jeans mass.

Besides individual brown dwarfs, it seems quite natural to suppose that --
much in the same way as it happens for ordinary stars -- also in this
case the fragmentation process should produce a substantial fraction
of binary brown dwarfs. For reasons that will become clear later, they
will be referred to as {\it primordial} binaries (we shall come
back to this issue in the next Sections). It is important to keep
in mind that the mass fraction of primordial binaries can be as large as
$50\%$ (Spitzer and Mathieu \cite{Mathieu}).

So, we are led to the conclusion that MACHOs consist of both individual
and binary brown dwarfs in this scenario (more about this, later).

However, we don't expect the fragmentation process to be able to convert the
whole gas in a PGC cloud into brown dwarfs.
For instance, standard stellar formation mechanisms lead to an upper 
limit of at most $40\%$ for the conversion efficiency (Scalo \cite{Scalo}).
Therefore, a fairly large amount of gas -- mostly $H_2$ -- should have been 
left over. What is its fate? At variance with the case of globular
clusters, strong stellar winds are now manifestly
absent, and so 
this gas should remain gravitationally bound in the dark clusters.
Actually, this conclusion is further supported by the following 
arguments (provided dark clusters comprise a consistent fraction of the
galactic dark matter). First, the gas cannot have diffused into the whole
halo, for otherwise it would have been heated by the gravitational
field to a virial temperature $\sim 10^6$ K, thereby becoming
observable in the X-ray band -- this option is ruled out by the
available upper limits (Dickey and Lockman \cite{DickeyLockman})
\footnote{Incidentally, the same argument also rules out a halo
primarily made of {\it unclustered} brown dwarfs (as well as white 
and M-dwarfs).}.
Second, the alternative possibility that the gas wholly collapsed
into the disk is also excluded, because the disk mass would then be of the
order of the inferred dark halo mass.
Now, the virial theorem entails that the temperature $T_{DG}$ of
a {\it diffuse} gas component inside a dark cluster is
\begin{equation}
T_{DG} \simeq 1.1 \left(\frac{M_{DC}}{M_{\odot}}\right)^{2/3} ~ {\rm K}~.
\label{eqno:R10}
\end{equation}
Accordingly, a large fraction of diffuse gas would
presumably give rise to an unobserved 
radio emission. Thus, we conclude that the amount of virialized 
diffuse gas
inside a dark cluster has to be low (it will henceforth be neglected).
This circumstance implies in turn that most of the leftover  gas should
be in the form of {\it self-gravitating} clouds clumped into the dark
clusters (since in this case the virial theorem applies to
individual clouds). As we shall see later, there are good reasons
to believe that the central temperature $T_m$ of the molecular clouds
in question should be very low, in fact close to that of the CBR.
Accordingly, the molecular cloud mass $M_m$ and median radius $r_m$ are 
related by the virial theorem as 
\begin{equation}
r_m \simeq 4.8\times 10^{-2} \left(\frac{M_m}{M_{\odot}}\right)~ {\rm pc}~. 
\label{eqno:R11}
\end{equation}
Presumably, the fraction of cluster dark matter in the form
of molecular clouds should be a function of the galactocentric distance $R$,
depending on the environmental conditions, like the UV flux and the 
collision rate for the PGC clouds.
 
Before proceeding further, an important issue should be faced. Given
the supposed existence of a large amount of gas in the dark clusters,
one would expect a star formation process to be presently operative.
However, things are not so simple. For -- under the above
environmental conditions -- only stars with mass smaller than the
cloud mass can be formed. Evidently, these stars are again either 
brown dwarfs or M-dwarfs. So, we see that undetected bright stars
should not form in the dark clusters to the extent that our
assumptions hold true. One might also wonder whether a sizable amount
of gas is eventually left over. As already pointed out, we argue that
this should be the case, for otherwise it would mean that the brown or
M-dwarf formation mechanism should be much more efficient than any
known star formation mechanism. Moreover, Gerhard and Silk
\cite{Gerhard} have shown that the cluster gravitational field can
stabilize the clouds against collapse.

Unfortunately, the lack of any observational information about
dark clusters would make any effort to understand their structure and
dynamics hopeless, were it not for some remarkable
insights that our unified treatment of globular and dark clusters
provides us.

In the first place, it looks quite natural to assume that also dark
clusters have a denser core surrounded by an extended spherical
halo. For simplicity, we shall suppose throughout that the core
density profile can be taken as constant.
Moreover,  it seems reasonable
to imagine (at least tentatively) that dark clusters have the same
average mass density as globular clusters. Hence, we obtain
\begin{equation}
r_{DC}\simeq 0.12 \left(\frac{M_{DC}}{M_{\odot}} \right)^{1/3}~{\rm pc},
\label{rdc}
\end{equation}
where $M_{DC}$ and $r_{DC}$ denote the mass and the median radius of a
dark cluster, respectively. In addition, dark clusters -- just like
globular clusters -- presumably stay for a long time in
a quasi-stationary phase, with an average central density 
$\rho_*(0)$ slightly lower than
$10^4~M_{\odot}$ pc$^{-3}$ (which is the observed average central
density for globular clusters).

As a further implication of the present model, we stress that -- at
variance with the case of globular clusters -- the mass spectrum 
of the dark clusters should be smooth, since the monotonic
decrease of the PGC cloud temperature fails to single out any
particular mass scale. 
As it will be shown in Sect. 3, dark clusters in the mass range
$3\times 10^2~M_{\odot}\ut < M_{DC}\ut < 10^6~M_{\odot}$ should
survive all disruptive effects, and so we shall restrict our attention
to such a mass range throughout.

As far as dark clusters are concerned, we have seen that the brown
dwarf mass is expected to lie in the range
$10^{-2}-10^{-1}~M_{\odot}$. For definiteness (and with an eye to
microlensing results), we imagine that all individual brown dwarfs have
the same mass $m\simeq 0.1~M_{\odot}$. So, binary brown
dwarfs are twice as heavy. As a consequence, the mass stratification
instability (Spitzer \cite{spitzer1969}) will drive them into the dark
cluster cores, which then tend to be composed chiefly by 
binaries. Furthermore, an average MACHO mass somewhat larger than 
$\simeq 0.1~M_{\odot}$ can naturally be accounted for.

Finally, let us consider molecular clouds. Since they also originate
from the above-mentioned fragmentation process,
we suppose (for definiteness) that they lie in the mass
range $10^{-3}~M_{\odot}\ut < M_m \ut < 10^{-1}~M_{\odot}$.
Correspondingly, eq. (\ref{eqno:R11}) entails
$4.8\times 10^{-5}~{\rm pc}\ut < r_m \ut < 4.8\times 10^{-3}~{\rm pc}$ and
$2.7\times 10^{10}~{\rm cm}^{-3}\ut > n_m \ut > 2.7\times 10^{6}~{\rm
cm}^{-3}$, respectively, where $n_m$ denotes the number density in
the clouds.

Before closing this Section, some comments are in order.
There is little doubt that the foregoing considerations
are qualitative in nature.
Nevertheless, they provide nontrivial insights into several questions that
arise in connection with the discovery of MACHOs.
Specifically,  a sharply peaked IMF in the range 
$10^{-2}-10^{-1}~M_{\odot}$ comes out naturally without having to
invoke any new physical process. This fact explains the
observed absence of a substantial amount of main-sequence stars inside the
dark clusters. 
Therefore, the observed absence of a large number of planet-like objects in the
halo (Renault et al. \cite{renault}) is automatically explained.
Furthermore, we can understand why brown dwarfs 
clumped into dark clusters copiously form in the outer halo, 
but neither in the inner halo nor
in the disk. Indeed,  the different stellar content of
these regions is here traced back to the different environments
in which the {\it same} star formation mechanism operates.
It goes without saying that various issues addressed above require further
investigations.

\section{Dynamical constraints on dark clusters}
As we have seen, MACHOs are clumped into dark clusters  when they form in the 
outer galactic halo. Still, the further fate of these clusters is quite 
unclear. For, they might either evaporate or drift towards the galactic centre.
In the latter case, encounters with globular clusters might have dramatic 
observational consequences and dynamical friction could drive too many MACHOs
into the galactic bulge. So, even if dark clusters are unseen, 
nontrivial constraints on their characteristic
parameters arise from the observed properties of our galaxy.
Moreover, in order to play any r\^ole as a candidate for dark matter,
MACHOs must have survived until the present in the outer part of the
galactic halo. Finally, it is important to know whether MACHOs are still
clumped into clusters today, especially because an improvement in the
statistics of microlensing observations permits to test this possibility
(Maoz \cite{maoz}, Metcalf and Silk \cite{ms}).

We remark that previous work on dynamical constraints on clusters of brown
dwarfs (Carr and Lacey \cite{lacey}, Carr \cite{bcarr}, Moore and Silk
\cite{moore}, Gerhard and Silk \cite{Gerhard}, Kerins \cite{kerins}) 
rests upon the hypothesis of an initial dark cluster 
distribution that extends inside the inner part of the Galaxy all the
way down to the centre. Hence,
a novel {\it ab initio} analysis is required within the present scenario.

We begin with the usual assumption that the halo dark matter density is 
modelled by an isothermal sphere \footnote{As mentioned in Sect. 1, this
model may not provide the best description of the Galaxy. 
However, it will
still be used in the present Section since otherwise the ensuing discussion
would become exceedingly complicated. Furthermore, it should be kept in
mind that the various uncertainties affecting the dark cluster properties
anyway make the following results reliable only as order-of-magnitude
estimates.} with density profile
\begin{equation}
\rho(R)=\rho_0~\frac{a^2 +R_{0}^2}{a^2 +R^2}
\label{density}
\end{equation}
where $a\simeq 5$ kpc is the halo core radius, $R_{0}\simeq 8.5$ kpc is our
galactocentric distance and $\rho_0 \simeq 6.5\times 10^{-25}$ g cm$^{-3}$
denotes the local dark matter density
corresponding to the one-dimensional velocity dispersion 
$\sigma \simeq 155$ km s$^{-1}$ of the halo.  Furthermore, we shall 
suppose for definiteness that the age of the Universe  is 
$t_0\simeq 10^{10}$ yr.
Our treatment of encounters rests upon the diffusion 
approximation and we will follow rather closely the analysis of Binney and
Tremaine \cite{binney}.

\subsection{Dynamical friction}
Dark clusters are subject to dynamical friction
as they orbit through the Galaxy, which makes them loose energy and therefore 
spiral in toward the galactic centre. Assuming (for illustrative purposes) 
that a dark cluster moves with velocity $v$ on a circular orbit of 
radius $R$, the drag brought about by the background density $\rho(R)$ is
given by
\begin{equation}
F(R)=-~0.43~\frac{4\pi G^2M_{DC}^2\ln\Lambda}{v^2}~\rho(R)
\label{friction}
\end{equation}
where $\ln \Lambda$ is the usual Coulomb logarithm, whose value in 
the present case is 
\begin{equation}
\ln\Lambda\simeq 
\ln\left(\frac{R_{typ} v^2}{GM_{DC}}\right)\simeq
24.3-\ln(M_{DC}/M_{\odot})~,
\end{equation}
with $R_{typ} \sim 20$ kpc,
$v\simeq \sqrt{2}\sigma$. Using eq. (\ref{density}), 
eq. (\ref{friction}) becomes
\begin{equation}
F(R)\simeq -~5.1\times 10^{14}\ln\Lambda\left(\frac{M_{DC}}{M_{\odot}}
\right)^2 \frac{1}{1+R_5^2}~{\rm g~cm~s^{-2}},
\end{equation}
where $R_5$ is in units of 5 kpc. 
Accordingly, the equations of motion entail that
a dark cluster originally at galactocentric distance $R$ will be
closer to the galactic centre today by the amount
\begin{equation}
\Delta R(R)\simeq 1.85\times 10^{-8}\left(\frac{M_{DC}}{M_{\odot}}
\right)\frac{24.3-\ln(M_{DC}/M_{\odot})}{R_5 + R_5^{-1}}
~{\rm kpc}.
\end{equation}
Keeping in mind that in our model $R > 10-20$ kpc and $M_{DC}\ut <
10^6~M_{\odot}$ (see later), we  see that 
$\Delta R\ut < 5.8\times 10^{-2}$ kpc.
Therefore, {\it dark clusters are still confined in the outer galactic halo}.
As a consequence, encounters between dark and globular clusters as well as
disk and bulge shocking of dark clusters are dynamically irrelevant, as long 
as they move on not too highly elongated orbits (in this way an effective
circularization of the orbits is achieved).

\subsection{Encounters between dark clusters}
Encounters between dark clusters may -- under the circumstances
to be analyzed below -- lead to their disruption. As an orientation, 
we notice that an estimate of the one-dimensional velocity dispersion
$\sigma_*$ of MACHOs and molecular clouds within a 
dark cluster is supplied by the virial theorem and reads
\begin{equation}
\sigma_*\simeq 6.9\times 10^{-2}\left(\frac{M_{DC}}{M_{\odot}}\right)^{1/3}
~{\rm km ~s^{-1}}~,
\label{sigmastella}
\end{equation}
where eq. (\ref{rdc}) has been used.
Because in the present scenario $M_{DC}\ut < 10^6~M_{\odot}$ (see later), 
we get $\sigma_*\ut < 6.9$ km s$^{-1}$. Therefore, the 
one-dimensional velocity dispersion of dark clusters - 
which we naturally suppose to be just 
$\sigma$ - is much larger than $\sigma_*$. Hence it makes sense to work 
within the impulse approximation, whose range of validity is 
established more precisely by the condition 
$M_{DC}\ll 10^{10}~M_{\odot}$,
which is evidently always met in our model.

In order to proceed further, we denote by $\Delta E$ the change of the
internal energy of a dark cluster in a single encounter.
Then (following Binney and Tremaine \cite{binney}) we find that encounters
with impact parameter $b$ in the range $b_{\rm min}\le b\le b_{\rm max}$
increase the cluster's energy at the rate
\begin{equation}
\dot{E}(R)\simeq\sqrt{\pi}~\frac{n_{DC}(R)}{\sigma^3}\int_0^{\infty}dv
v^3 e^{-v^2/4\sigma^2}\int_{b_{\rm min}}^{b_{\rm max}}db~ b~ \Delta E
\label{dote}
\end{equation}
where $v$ and $n_{DC}(R)$ are the cluster velocity and number density (in
the halo), respectively. We let $\gamma$ stand for the fraction of halo dark
matter in the form of dark clusters, so that we have
$n_{DC}(R)=\gamma \rho(R)/M_{DC}$, with $\rho(R)$ given
by eq. (\ref{density}). Accordingly, eq. (\ref{dote}) becomes
\begin{equation}
\dot{E}(R)\simeq\frac{1}{2\sqrt{\pi}}~\frac{\gamma}{\sigma GM_{DC}}
~\frac{1}{R^2 + a^2}\int_0^{\infty}dv
v^3 e^{-v^2/4\sigma^2}\int_{b_{\rm min}}^{b_{\rm max}}db~ b~\Delta E~.
\label{dote1}
\end{equation}
Now, a natural definition of the time required by encounters
to dissolve a cluster is provided by $t_d(R)=E_{\rm bind}/\dot{E}(R)$,
where the binding energy $E_{\rm bind}$ is expressed in terms 
of the cluster properties as $E_{\rm bind}\simeq 0.2~ GM_{DC}^2/r_{DC}$.

\subsubsection{Distant encounters}
Let us consider first distant encounters. Correspondingly,
$\Delta E$ is to be evaluated in the tidal approximation (Spitzer 
\cite{spitzer1})
and reads presently 
\begin{equation}
\Delta E\simeq \frac{4G^2 M_{DC}^3 r_{DC}^2}{3b^4 v^2}~.
\end{equation}
We insert this quantity into eq. (\ref{dote1}), neglecting the term
$(b_{\rm min}/b_{\rm max})^2$ in the ensuing expression.
Experience with a similar treatment for globular clusters suggests to 
choose $b_{\rm min}\simeq r_{DC}$. 
Correspondingly, we find
\begin{equation}
t_d(R)\simeq \frac{4.7\times 10^{11}}{\gamma}
\left(\frac{M_{\odot}}{M_{DC}}\right)^{1/3}(R_5^2 +1)~{\rm yr},
\end{equation}
where use of eq. (\ref{rdc}) has been made. Assuming $R>10-20$ kpc and 
keeping in mind that $\gamma\le 1$ and $M_{DC}\ut < 10^6~M_{\odot}$
(see later), we get that for all the dark clusters 
in question $t_d(R)$ exceeds the age of the Universe.

\subsubsection{Close encounters}
In order to deal with close encounters, dark clusters have to be regarded
as extended objects.
As in the case of globular clusters, this task is most simply accomplished by 
modelling the dark clusters by means of a Plummer potential with core radius 
$\alpha$. Correspondingly, $\Delta E$ is found to be
\begin{equation}
\Delta E\simeq \frac{G^2 M_{DC}^3}{3\alpha^2 v^2}~.
\end{equation}
As before, we insert this quantity into eq. (\ref{dote1}), assuming now
$b_{\rm min}\simeq 0$ and $b_{\rm max}\simeq r_{DC}$. 
In this way, we obtain
\begin{equation}
t_d(R)\simeq\frac{2.4\times 10^{11}}{\gamma}~
\left(\frac{\alpha}{r_{DC}}\right)^2 
\left(\frac{{\rm pc}}{r_{DC}}\right)
(R_5^2 +1)~{\rm yr}~.
\label{td}
\end{equation}
We proceed by recalling that
\begin{equation} 
\alpha=\left(\frac{3M_{DC}}{4\pi \rho_*(0)} \right)^{1/3}~,
\end{equation}
where $\rho_*(r)$ denotes the dark cluster mass density. Taking 
$\rho_*(0)\simeq 10^4~M_{\odot}$ pc$^{-3}$ (as for globular clusters 
today) and using eq. (\ref{rdc}), we can rewrite eq. (\ref{td}) as
\begin{equation}
t_d(R)\simeq \frac{10^{11}}{\gamma}
\left(\frac{M_{\odot}}{M_{DC}}\right)^{1/3}(R_5^2 +1)~{\rm yr}.
\end{equation}
Assuming $R> 10-20$ kpc and remembering that $\gamma\le 1$, we are led 
to the conclusion 
that {\it dark clusters are not disrupted by close encounters}
provided $M_{DC}\ut < 10^6~M_{\odot}$. \footnote{It should hardly come as a 
surprise that close encounters yield a more stringent bound on $M_{DC}$ 
than distant encounters.}

\subsection{Evaporation}
Various dynamical effects conspire to make dark clusters evaporate within
a finite time. Relaxation via gravitational two-body encounters leads to the 
escape of
MACHOs approaching the unbound tail of the cluster velocity distribution.
Tidal truncation due to the galactic gravitational field enhances this process.
A more substantial effect is caused by the gravothermal instability, when
the inner part of the dark clusters contracts (core collapse)
and the envelope expands.

Below, we shall address these issues separately. As is well known, a 
key-r\^ole in such an analysis is 
played by the relaxation time 
\begin{equation}
t_{\rm relax}(r)=0.34~\frac{\sigma_*^3}{G^2 m \rho_*(r)\ln(0.4N)}
\label{relaxation}
\end{equation}
where $N$ is the number of MACHOs
per cluster. 
As in the case of globular clusters, $\rho_*(r)$ is expected to vary by 
various orders of magnitude
in different regions of a single dark cluster -- this dependence  
obviously shows up in $t_{\rm relax}(r)$. Therefore, for reference purposes, it
is often more convenient to characterize a dark cluster by a single value of 
the relaxation time. This goal is achieved by introducing the median relaxation
time (Spitzer and Hart \cite{sh})
\begin{equation}
t_{\rm rh}=\frac{6.5\times 10^8}{\ln(0.4N)}
\left(\frac{M_{DC}}{10^5~M_{\odot}}\right)^{1/2}
\left(\frac{M_{\odot}}{m}\right)
\left(\frac{r_{DC}}{{\rm pc}}\right)^{3/2}~{\rm yr}~.
\label{trh}
\end{equation}
Explicitly, using eq. (\ref{rdc}) together with eq. (\ref{sigmastella}) 
and $m\simeq 0.1~M_{\odot}$, eq. ({\ref{relaxation}) takes the form
\footnote{For simplicity, we neglect here the fact that
binaries have a mass larger than individual brown dwarfs.
Furthermore, given the logarithmic $N$-dependence, we can safely take $N\sim 
M_{DC}/m\sim10~ (M_{DC}/M_{\odot})$.}
\begin{equation}
t_{\rm relax}(r)\simeq 5\times 10^7
\left(\frac{M_{DC}}{M_{\odot}}\right)
\left(\frac{M_{\odot}~{\rm pc^{-3}}}{\rho_*(r)}\right)
\frac{1}{1.4+\ln(M_{DC}/M_{\odot})}~{\rm yr}~.
\end{equation}
In the same fashion, eq. (\ref{trh}) becomes
\begin{equation}
t_{\rm rh}\simeq 8\times 10^5
\left(\frac{M_{DC}}{M_{\odot}}\right)
\frac{1}{1.4+\ln(M_{DC}/M_{\odot})}~  {\rm yr}.
\end{equation}

\subsubsection{Spontaneous evaporation} 
As is well known, any stellar association evaporates spontaneously 
\footnote{We like to call in this
way the evaporation process which is neither induced by external perturbations
nor specific of the gravitational interactions.}
within
a finite time owing to relaxation via gravitational two-body encounters. 
Specifically, a single close encounter between two MACHOs can leave one
of them with a speed larger than the local escape velocity. So, the 
MACHO under consideration gets ejected from the dark cluster. 
We find for the ejection time (H\'enon \cite{henon}), 
\begin{equation}
t_{\rm ej}\simeq 1.1 \times 10^3~ {\rm ln}(0.4 N)~ t_{rh}
\simeq 9\times 10^{8}
\left(\frac{M_{DC}}{M_{\odot}}\right) ~{\rm yr}.
\label{tej}
\end{equation}
Alternatively, several
more distant weaker encounters can gradually increase the energy of
a given MACHO, until a further weak encounter is sufficient to make
it escape from the cluster. In this case, the evaporation time
turns out to be (Spitzer and Thuan \cite{st})
\begin{equation}
t_{\rm evap}\simeq 300~ t_{\rm rh}\simeq 2.4 \times 10^8
\left(\frac{M_{DC}}{M_{\odot}}\right)
\frac{1}{1.4+\ln(M_{DC}/M_{\odot})}~  {\rm yr}.
\label{tevap}
\end{equation}
Since $t_{\rm ej}$ is anyway longer than $t_{\rm evap}$, we shall focus our 
attention on the latter quantity. 
Hence, by demanding that $t_{\rm evap}$ should exceed the age of the 
Universe we conclude that {\it dark clusters with} $M_{DC}\ut >
3\times 10^2~M_{\odot}$  {\it are not yet evaporated}.

\subsubsection{Tidal Perturbations}
Dark clusters -- just like globular clusters -- are tidally disrupted by 
the galactic gravitational field unless $r_{DC}$ is smaller
than their tidal radius.
So, the survival condition reads 
\begin{equation}
r_{DC}<\left(\frac{M_{DC}}{3M_G(R)} \right)^{1/3} R~,
\label{rdcnew}
\end{equation}
where $R$ should be understood here as the perigalactic distance of
the dark cluster and $M_G(R)$ denotes the Galaxy mass inside $R$. From 
eq. (\ref{density}) we obtain
\begin{equation}
M_G(R)\simeq 5.5\times 10^{10}~
R_5 (1- R_5^{-1}\arctan R_5)~M_{\odot}~,
\end{equation}
and - on account of eq. (\ref{rdc}) - eq. (\ref{rdcnew}) becomes
\begin{equation}
R_5[1- R_5^{-1}\arctan R_5]^{-1/2}>0.047~,
\end{equation}
which is always satisfied for $R>10-20$ kpc. Thus, the dark clusters under 
consideration are not tidally disrupted by the galactic gravitational field.

\subsubsection{Core collapse}
An important r\^ole in the considerations to follow is played by core collapse.
It is by now well-established that the initial stage of this process is 
triggered by evaporation, which leads to the shrinking of the core as a 
consequence of energy conservation. Numerical Fokker-Planck
studies of the early phase of 
core collapse have shown that the dynamics of the cluster is correctly
described by a sequence of King models (Cohn \cite{ch}). 
However, once the cluster density reaches a certain critical value, 
core collapse gets dramatically accelerated by the gravothermal instability
(Antonov \cite{antonov}, Lynden-Bell and Wood \cite{lw}). 
Indeed, the negative specific heat of the core implies that the internal 
velocity dispersion $\sigma_*$ increases -- thereby enhancing evaporation --
as the average kinetic energy decreases by evaporation itself. Moreover,
the unbalanced gravitational energy makes the core contract, and so its
density rises by several 
orders of magnitude in a runaway manner.
Numerical simulations show that the central velocity dispersion and 
the number of stars in the core $N_*$ scale as
\begin{equation}
\sigma_{*} ~\sim~ \rho_*(0)^{0.05}~,
\label{a37}
\end{equation}
\begin{equation}
N_* ~\sim ~\rho_*(0)^{-0.36}~, \label{38}
\end{equation}
respectively (Binney and Tremaine \cite{binney}).
Incidentally, the somewhat surprising slow rise of $\sigma_*$ in 
eq. (\ref{a37}) is due to the large mass-loss from the core, as it
follows from eq. (\ref{38}).

When does the gravothermal instability show up? Unfortunately, a
clear-cut answer does not exist, since the corresponding time 
$t_{GI}$ depends on how clusters  are modelled as well as on 
their concentration (Quinlan \cite{quinlan}).
Manifestly, the lack of observational data about dark clusters
makes  a precise determination of $t_{GI}$ impossible. So, the best 
we can do is to suppose that dark clusters behave just like
globular clusters as far as core collapse is concerned. 
In this way, we are lead  to the order-of-magnitude estimate 
(Binney and Tremaine \cite{binney}) (more about this, later)
\begin{equation}
t_{GI}\simeq 3~t_{rh}\simeq 2.4\times 10^6
\left(\frac{M_{DC}}{M_{\odot}}\right)
\frac{1}{1.4+\ln(M_{DC}/M_{\odot})}~  {\rm yr}~.
\label{a39}
\end{equation}
Comparing now $t_{GI}$ with the age of the Universe, we conclude
that {\it the dark clusters with} $M_{DC}\ut < 5\times 10^4~M_{\odot}$
{\it are expected to have started core collapse}.

As we said, the  central density grows dramatically during the second
stage of core collapse, and so the central relaxation time gets
shorter and shorter. Detailed studies of the gravothermal instability
have shown that -- if nothing opposes to collapse -- the time needed
to complete core collapse $t_{\rm coll}$ starting from an arbitrary time $t$
goes like $t_{\rm relax}(0)$, with the latter quantity computed for the
particular value taken by $\rho_*(0)$ at $t$. More specifically,
computer simulations of globular cluster dynamics entail
$t_{\rm coll}\simeq 330 ~t_{\rm relax}(0)$ (Cohn \cite{ch}).

Because of the huge increase in the central density, close two-body
encounters lead to the formation of bound binary systems by converting
enough kinetic energy into internal energy (tidal capture).
As we will see in Section 4, binary brown dwarfs are
produced in this way in the dark cluster cores during the early phase of core
collapse. These binaries - which will be referred to 
as {\it tidally-captured} binaries and happen to be very hard - play a
crucial r\^{o}le in this context, since they ultimately stop and
reverse the collapse. Schematically, the argument goes as
follows. Owing to the fact that hard binaries necessarily get harder
in collisions with individual stars (Heggie \cite{heggie}),
the internal binding energy released
by a binary is transformed into kinetic energy of both the star and
the binary. Actually, the exchanged energy is so large that they both
leave the cluster. However -- at variance with evaporation -- the
kinetic energy (per unit mass) of the cluster is now unaffected, 
whereas mass ejection obviously increases the potential energy.
That is, the binding energy given up by the binaries ultimately
becomes gravitational energy of the core. As a result of the
unbalanced kinetic energy, the core starts expanding. Moreover, 
because of the negative specific heat the increased potential energy
makes $\sigma_{*}$ decrease, thereby slowing down mass ejection. 
In this manner, core collapse gets halted and reversed (Spitzer
\cite{spitzer1}). 

As a matter of fact, the presence of binaries in appreciable amount
can also modify to some extent the standard scenario of core collapse as
outlined above. Indeed, numerical Fokker-Planck simulations 
have shown that in this case the 
collapse is also driven by the mass stratification instability. As a
consequence, the collapse proceeds faster and starts before than in the 
few-binary case (the latter point makes eq.~(\ref{a39}) more 
plausible than it might appear at first sight). This phenomenon is found to 
occur both for tidally-captured binaries (Statler, Ostriker and 
Cohn \cite{soc}) and for primordial binaries
(Spitzer and Mathieu \cite{Mathieu}).

\subsection{Discussion}
What the above analysis shows is that dark clusters within the mass range
\newline
$3\times 10^2~M_{\odot} \ut < M_{DC} \ut < 10^6~M_{\odot}$ should have survived
all disruptive effects arising from gravitational perturbations and are 
nowadays  expected to populate the outer part of the
galactic halo. In addition, the clusters with 
$3\times 10^2~M_{\odot} \ut < M_{DC} \ut < 5\times 10^4~M_{\odot}$ should 
undergo core collapse. 
Unfortunately, it is practically impossible to predict the further fate 
of those dark clusters which are in the post-collapse phase today.
{\it A priori} it seems natural to imagine that bounce and
subsequent reexpansion should follow core collapse (Cohn and Hut \cite{chh},
Heggie and Ramamani \cite{hr}).
And perhaps also a whole 
series of core contractions and expansions can take place, giving rise to 
the so-called gravothermal oscillations (Bettwieser and Sugimoto \cite{bs}). 
However, this conclusion crucially depends on the 
unknown model which correctly describes dark clusters. For instance,
in the case of tidally-truncated models the cluster gets completely destroyed
within a finite time (Stodolkiewicz \cite{stodol}, Ostriker, Statler and
Lee \cite{osl}). Moreover, what certainly happens in either case is that 
the number of MACHOs in the core monotonically decreases with time.  
So, an unclustered MACHO population is expected to coexist with dark 
clusters in the outer  galactic halo (unless all dark clusters have
$M_{DC}\ut > 5\times 10^4~M_{\odot}$) -- detection of unclustered MACHOs
would therefore not rule out the present scenario.

\section{MACHOs as binary brown dwarfs}
As  already pointed out, it seems natural to suppose that a  
fraction of primordial binary brown dwarfs -- possibly as large as $50\%$ 
in mass -- should form along with individual
brown dwarfs as a result of the fragmentation process of the
PGC clouds. Subsequently, because of the mass stratification instability
primordial binaries will concentrate inside the dark cluster cores, 
which are therefore expected to be chiefly composed by binaries and
molecular clouds. In addition -- as far as dark clusters with 
$M_{DC} \ut < 5 \times 10^4 M_{\odot}$ are concerned --
a population of tidally-captured binary brown dwarfs 
ought to form in the dark cluster cores owing to the 
increased central density caused by  core collapse. 
Thus, a large fraction of binaries should be present inside the dark cluster
cores at a late stage of their evolution. Below, we will
try to make the discussion of this issue as 
much quantitative as possible.

\subsection{Survival and hardness of binary brown dwarfs}
The first question that has to be addressed is whether a binary brown
dwarf -- produced by whatever mechanism long ago -- survives up until
the present. To this end, we recall that a binary system is {\it hard}
when its binding energy exceeds the kinetic energy of field stars
(otherwise it is {\it soft}). 
In the present case, binary brown dwarfs
happen to be hard when their orbital radius $a$ 
obeys the following constraint
\begin{equation}
a \ut <1.4\times 10^{12}\left(\frac{M_{\odot}}{M_{DC}}\right)^{2/3}~{\rm km}~.
\label{a1}
\end{equation}

As is well known, soft binaries always get softer whereas hard
binaries always get harder because of encounters with individual stars
(Heggie \cite{heggie}). So -- were individual-binary encounters the
only relevant process -- we would conclude that hard binary brown dwarfs
should indeed survive.
However, also binary-binary encounters play an important r\^ole in the
dark cluster cores, where binaries are expected to be
far more abundant than individual brown dwarfs.
Now, in the latter process one of the two binaries gets often
disrupted (this cannot happen for both binaries -- given that they are hard --
while fly bys are rather infrequent), thereby leading 
to the depletion of the binary population. 
We will address this effect in Sect. 4.3, where we shall
find that for realistic values of the dark cluster parameters the binary 
break up does not take place.

\subsection{Tidally-captured binary brown dwarfs}
As far as globular clusters are concerned, it is nowadays well known
that the most efficient mechanism for late binary formation is
dissipative tidal capture in the core (Fabian, Pringle and Rees
\cite{fabian}, Press and Teukolsky \cite{press}, 
Lee and Ostriker \cite{lee}). 
Hence, we expect a similar situation to occur for dark clusters as well.

Let us now analyze this phenomenon in a quantitative fashion.
As a first step, we observe that the radius $r_*$ of
a brown dwarf of mass $\simeq 0.1 M_{\odot}$ is 
$r_*\simeq 0.7\times 10^5$ km (Saumon et al. \cite{saumon}). Furthermore, the
Safronov number (Binney and Tremaine \cite{binney}) is
\begin{equation}
\Theta=\frac{Gm}{2~\sigma_*^2 ~r_*}\simeq 2\times 10^7\left(
\frac{M_{\odot}}{M_{DC}} \right)^{2/3}~,
\end{equation}
which turns out to be anyway much larger than one. 
Within this approximation, the time for brown dwarf tidal capture can be 
written as (Lee and Ostriker \cite{lee})
\begin{equation}
t_{\rm tid}\simeq 10^{12}
\left(\frac{10^5~ {\rm pc}^{-3}}{n_{IBD}(0)}\right)
\left(\frac{\sigma_*}{100~ {\rm km~s^{-1}}} \right)^{1.2}
\left(\frac{R_{\odot}}{r_*} \right)^{0.9}
\left(\frac{M_{\odot}}{m} \right)^{1.1}~{\rm yr},
\label{bdtidalcapture}
\end{equation}
where $n_{IBD}(0)$ is the number density of individual
brown dwarfs in the core
\footnote{Although eq. (\ref{bdtidalcapture}) 
has been derived for main-sequence
stars, it looks plausible that it applies also to brown dwarfs, since it
is not very sensitive to the particular stellar model (Lee and Ostriker
\cite{lee}).}.
Obviously, we have $n_{IBD}(0)\simeq f_{IBD}~\rho_*(0)/m$, 
with $f_{IBD}$ denoting the mass fraction of individual brown dwarfs in 
the core.
From eq. (\ref{a37}) we see that $\sigma_*$ increases very slightly, and so
core collapse effects on $\sigma_*$ can safely be neglected. 
Accordingly, we can rewrite eq. (\ref{bdtidalcapture}) as
\begin{equation}
t_{\rm tid}\simeq \frac{1.6\times 10^{14}}{f_{IBD}}
\left(\frac{M_{\odot}~{\rm pc^{-3}}}{\rho_*(0)}\right)
\left(\frac{M_{DC}}{M_{\odot}}\right)^{0.4}~{\rm yr}
\label{bdtidal}
\end{equation}
where eq. (\ref{sigmastella}) has been used.
Comparing now $t_{\rm tid}$ with the age of the Universe, we get that
practically all individual brown dwarfs in the core are tidally captured into
binaries provided
\footnote{More precisely, we are demanding that the rate for tidal 
capture times the age of the Universe should exceed one. Notice that
the former quantity is one-half of $t_{\rm tid}^{-1}$, and so we
require that $2t_{\rm tid}$ should be smaller than $10^{10}$ yr (we are
actually following the same procedure used by Press and Teukolsky 
\cite{press}).}
\begin{equation}
\rho_*(0)> \frac{3.2 \times 10^4}{f_{IBD}} \left(\frac{M_{DC}}{M_{\odot}}
\right)^{0.4} ~M_{\odot}~{\rm pc}^{-3}~. 
\label{rho}
\end{equation}
According to the above assumptions, we expect $\rho_*(0) \simeq
10^4~M_{\odot}~{\rm pc}^{-3}$ just before core collapse.
Therefore, we see that tidal capture
requires an increase of $\rho_*(0)$ by a factor in 
the range
$31/f_{IBD} - 242/f_{IBD}$, corresponding to $M_{DC}$
in the range $3 \times 10^2~M_{\odot} - 5\times 10^4
~M_{\odot}$. Thus, the formation of tidally-captured
binaries would occur during the early phase 
of core collapse (the same conclusion was reached in a different way by 
Statler, Ostriker and Cohn \cite{soc} for globular clusters).
However, this conclusion depends on the fractional (mass) abundance
$f_{PB}$ of primordial binaries in the core, since $f_{IBD}$ necessarily
gets small for large $f_{PB}$.

Next, we compute the (average) orbital radius of tidally-captured
binary brown dwarfs following the procedure outlined by Statler,
Ostriker and Cohn \cite{soc}. Correspondingly, we find 
$a\simeq 2.5\times 10^5$ km (this value is practically independent 
of $M_{DC}$). As a consequence, we see that 
they are so hard that
condition (\ref{a1}) is always abundantly met.

Let us finally try to estimate the fractional abundance of tidally-captured
binary brown dwarfs in the dark cluster cores soon after their 
formation, namely when 
inequality (\ref{rho}) just starts to be satisfied (their total number 
at this stage will be
denoted by $N_{TCB}$). Thanks to eq. (\ref{38}) we easily get
\begin{equation}
N_{TCB}\simeq 3.3 f_{IBD}^{1.36}\left(\frac{M_{DC}}{M_{\odot}}\right)^{0.86}
\left(\frac{M_c}{M_{\odot}}\right)~,
\label{ntcb}
\end{equation}
where $M_c$ denotes the core mass just before core collapse (corresponding to
$\rho_*(0)\simeq 10^4~M_{\odot}~{\rm pc}^{-3}$).
Denoting further by $N_{IBD}^{\rm tot}$ the total number of individual
brown dwarfs in a dark cluster before core collapse, we find
\begin{equation}
\frac{N_{TCB}}{N_{IBD}^{\rm tot}}\simeq 0.33 f_{IBD}^{0.36}\left(
\frac{M_{\odot}}{M_{DC}}\right)^{0.14}
\left(\frac{M_c}{M_{DC}}\right)~.
\label{ntcb*}
\end{equation}
Realistically, even in the extreme case of a fully baryonic halo we expect
$f_{IBD}\ut < 0.3$ (of course, a large $f_{PB}$ would imply a $f_{IBD}$
considerably smaller than that). 
Moreover, $M_c/M_{DC}$ sensitively depends on how the core 
is defined, but the analogy with globular clusters
entails that it should anyway be less than $20\%$. 
Accordingly, we conclude that the fractional abundance of tidally-captured
binary brown dwarfs should not exceed $1\%$ (again in remarkable agreement
with the result of Statler, Ostriker and Cohn \cite{soc} for globular
clusters), and so they are irrelevant from the observational point of view.

\subsection{Primordial binary brown dwarfs}
Primordial binaries are a very different story. For, not only are they
expected to be much more abundant than tidally-captured binaries, but
in addition they are presumably much less hard, since all values for
their orbital radius consistent with condition (\ref{a1}) are in
principle to be contemplated. So, hardening effects ought to play a crucial
r\^ole for primordial binaries (as it can be guessed on intuitive grounds,
effects of this kind turn out to be totally negligible for 
tidally-captured binaries -- this will be demonstrated later on).

\subsubsection{Collisional hardening} 

Let us begin by considering 
collisional hardening, namely the process whereby hard
binaries get harder in encounters with individual brown dwarfs. We recall
that the associated average hardening rate (Spitzer
\& Mathieu 1980) reads presently
\begin{equation}
\dot E~ \simeq~ -~2.8~ \frac{G^2 m^3~ n_{IBD}(0)}{\sigma_*}~.
\label{11}
\end{equation}
On account of the definition of $n_{IBD}(0)$ and of eq. (\ref{sigmastella}),
eq. (\ref{11})
becomes
\begin{equation}
\dot E~  \simeq  ~ -~1.7 \times 10^{32} f_{IBD}
\left(\frac{M_{\odot}}{M_{DC}} \right)^{1/3} 
\left(\frac{\rho_*(0)}{M_{\odot} ~{\rm pc}^{-3}} \right) 
~~{\rm erg}~{\rm yr}^{-1}~.
\label{edot}
\end{equation}
Observe that a characteristic feature of collisional hardening is that
$\dot E$ is independent of hardness, and so it is time-independent.

As is well known, the internal energy of a binary is 
$E=-Gm^2/2a$, which yields
\begin{equation}
\dot E = \frac{G m^2}{2 a^2} ~\dot a~.
\label{aa}
\end{equation}
Hence, by combining eqs. (\ref{edot}) and (\ref{aa}) together and
integrating the ensuing expression we get 
\begin{equation}
\frac{{\rm km}}{a_2} \simeq \frac{{\rm km}}{a_1} + 1.3 \times 10^{-20}
f_{IBD} \left(\frac{M_{\odot}}{M_{DC}}\right)^{1/3}
\left(\frac{\rho_{*}(0)}{M_{\odot}~ {\rm pc}^{-3}}\right)
\left( \frac{t_{21}}{{\rm yr}}\right)~,
\label{a11}
\end{equation}
where $a_1$ stands for the initial orbital radius, whereas
$a_2$ denotes the orbital radius after a time $t_{21}$.

Assuming momentarily that no other hardening mechanism is
operative and taking $t_{21}$ equal to the age of the Universe, we find 
that the present orbital radius of a binary brown dwarf is given by
\begin{equation}
\frac{{\rm km}}{a_2} \simeq \frac{{\rm km}}{a_1} + 
1.3 \times 10^{-10} f_{IBD}
\left( \frac{M_{\odot}}{M_{DC}} \right)^{1/3} \left( \frac{\rho_{*}(0)}
{M_{\odot}~ {\rm pc^{-3}}} \right)~.
\label{a2}
\end{equation}
Of course, collisional hardening works to the extent that $a_2$ becomes
considerably smaller than $a_1$. Correspondingly, we see from eq. (\ref{a2}) 
that this is indeed the case provided
\begin{equation}
a_1 \ut > 8 \times 10^9 f_{IBD}^{-1} \left(\frac{M_{DC}}{M_{\odot}}
\right)^{1/3} \left(\frac{M_{\odot}~ {\rm pc}^{-3}}{\rho_{*}(0)} \right)~
{\rm km} ~.
\label{a3}
\end{equation}
Thanks to eq. (\ref{a2}), eq. (\ref{a3}) yields in turn
\begin{equation}
a_2 \simeq 8 \times 10^9 f_{IBD}^{-1} \left(\frac{M_{DC}}{M_{\odot}}
\right)^{1/3} \left(\frac{M_{\odot}~ {\rm pc}^{-3}}{\rho_{*}(0)} \right)~
{\rm km} ~.
\label{a4}
\end{equation} 

Physically, the emerging picture is as follows. Only those binaries 
whose initial orbital radius obeys condition (\ref{a3}) undergo collisional
hardening, and their present orbital radius turns out to be almost
{\it independent} of the initial value. We can make the present discussion 
somewhat more specific by noticing that our assumptions strongly suggest
$\rho_{*}(0) \ut < 10^4 M_{\odot}~ {\rm pc}^{-3}$, in which case both
eq. (\ref{a3}) and eq. (\ref{a4}) acquire the form
\begin{equation}
a_{1,2} \ut > 8 \times 10^5 f_{IBD}^{-1} \left( \frac{M_{DC}}{M_{\odot}}
\right)^{1/3} ~{\rm km}~.
\label{a5}
\end{equation}

Evidently, very hard primordial binaries -- which fail to meet
condition (\ref{a5}) -- do not suffer collisional hardening, and the same
is true for tidally-captured binaries.

\subsubsection{Frictional hardening}
As we are going to show, the presence of molecular 
clouds in the dark cluster cores
-- which  is indeed the most characteristic feature of the model in
question -- provides
a novel hardening mechanism for binary brown dwarfs.
Basically, this is brought about by
dynamical friction  on molecular clouds.

It is not difficult to extend the standard treatment of dynamical friction
(Binney \& Tremaine \cite{binney}) to the relative motion of the
brown dwarfs in a binary system which moves inside a molecular cloud.
For simplicity, we assume that molecular clouds have a constant density
profile $\rho_m$. In the case of a circular orbit \footnote{Indeed, the
circularization of the orbit is achieved by tidal effects after a
few periastron passages (Zahn 1987).}, 
the equations of motion imply that the time $t^{(0)}_{21}$ 
needed to reduce the
orbital radius $a$ of a binary which moves 
{\it all the time} inside molecular clouds from $a_1$ down to $a_2$ is
\begin{equation}
t^{(0)}_{21} \simeq 0.17\left(\frac{m}{G}\right)^{1/2}
\frac{1}{\rho_m\ln\Lambda}~(a_{2}^{-3/2}-a_{1}^{-3/2})~,
\label{t21}
\end{equation}
where the Coulomb logarithm reads
\begin{equation}
\ln\Lambda\simeq \ln ({r_m v_c^2}/{Gm})\simeq\ln ({r_m}/{a_1})~,
\label{lambda}
\end{equation}
with $v_c$ denoting the circular velocity (approximately
given by Kepler's third law). Manifestly, the diffusion approximation -- 
upon which the present treatment is based -- requires that the orbital 
radius of a binary should always be smaller than the median radius of a 
cloud.
As we are concerned henceforth with hard
binaries, $a_1$ has to obey condition (\ref{a1}). On the other hand, $a_1$ is
the larger value for the orbital radius in eq. (\ref{t21}). So, we
shall take for definiteness -- in the Coulomb logarithm only -- 
$a_1\simeq 1.4\times 10^{12}(M_{\odot}/M_{DC})^{2/3}$ km. In addition,
from eq. (\ref{eqno:R11}) we have 
\begin{equation}
\rho_m\simeq 2.5 ~\left(\frac{\rm pc}{r_m}\right)^2~~M_{\odot}~{\rm pc}^{-3}~. 
\label{56}
\end{equation}
Hence, putting everything together we obtain
\begin{equation}
\left(\frac{{\rm km}}{a_2}\right)^{3/2} \simeq
\left( \frac{{\rm km}}{a_1}\right)^{3/2} + 2 \times 10^{-26}~ \Xi^{-1}
\left(\frac{{\rm pc}}{r_m}\right)^2 \left(\frac{t^{(0)}_{21}}{\rm yr}
\right)~,
\label{a6}
\end{equation}
having set
\begin{equation}
\Xi\equiv [3+\ln({r_m}/{{\rm pc}})+0.7\ln({M_{DC}}/{M_{\odot}})]^{-1}~.
\label{15}
\end{equation}
Specifically, the diffusion approximation demands $\Xi >0$, which yields in 
turn
\begin{equation}
r_m >5\times 10^{-2}\left(\frac{M_{\odot}}{M_{DC}}\right)^{0.7}~~{\rm pc}~.
\label{rrmm}
\end{equation}
Observe that for $M_{DC}\ut < 2.1\times 
10^4~M_{\odot}$ this constraint restricts the range of allowed values of
$r_m$ as stated in Sect. 2.2.

Were the dark clusters completely filled by clouds, eq. (\ref{a6}) 
would be the final result. 
However, the distribution of the clouds is lumpy. 
So, if we want to know the orbital radius $a_2$ after a time $t_{21}$
we have to proceed as follows. First, we should compute the fraction
$t^{(0)}_{21}$ of the time interval in question $t_{21}$
spent by a binary inside the clouds. Next, we have to
re-express $t^{(0)}_{21}$ in eq. (\ref{a6}) in terms of $t_{21}$.
This goal will be achieved by the procedure outlined below.

Keeping in mind that both the 
clouds and the binaries have
average velocity $v\simeq \sqrt{3}\sigma_*$ (for simplicity, we
neglect the equipartition of kinetic energy of the binaries)
it follows that the time needed by a binary to cross a single cloud is
\begin{equation}
t_m\simeq \frac{r_m}{\sqrt{2}~v}\simeq
5.6\times 10^6\left(\frac{r_m}{{\rm pc}}\right)
\left(\frac{M_{\odot}}{M_{DC}}\right)^{1/3}~{\rm yr}~.
\label{16}
\end{equation}
As an indication, we notice that for $r_m\simeq 10^{-3}$ pc 
($M_m \simeq 2 \times 10^{-2} ~M_{\odot}$) and
$M_{DC}\simeq 10^5~M_{\odot}$ we find 
$t_m\simeq 1.2\times 10^2$ yr. Therefore, frictional  
hardening involves many clouds.
Specifically, during the time $t_{21}$ the number of clouds crossed
by a binary is evidently
\begin{equation}
N_m\simeq\displaystyle{\frac{t^{(0)}_{21}}{t_m}}
\simeq 1.8\times 10^{-7}
\left(\displaystyle{\frac{{\rm pc}}{r_m}}\right)
\left(\frac{M_{DC}}{M_{\odot}}\right)^{1/3} \left(\frac{t^{(0)}_{21}}
{{\rm yr}} \right)~.
\label{tildeN1}
\end{equation}
Let us now ask how many crossings of the core are necessary
for a binary to traverse $N_m$ clouds.
To this end, we proceed to estimate the number of clouds $N_c$ 
encountered during {\it one} crossing of the core. 
Describing the dark clusters by a King model, we can identify the core
radius with the King radius. Evidently, the cross-section for binary-cloud
encounters is $\pi r_m^2$ and so we have
\begin{equation}
N_c\simeq
\left(\frac{9\sigma_*^2}{4\pi G\rho_*(0)}\right)^{1/2} n_{CL}(0)~ \pi~r_m^2~,
\label{tildeN2}
\end{equation}
with $n_{CL}(0)$ 
denoting the cloud number density in the core.
Thanks to eq. (\ref{eqno:R11}), we can write
\begin{equation}
n_{CL}(0)  \simeq  f_{CL}\frac{\rho_*(0)}{M_m} 
\simeq 
4.8\times 10^{-2}f_{CL}\left(\frac{{\rm pc}}{r_m}\right)
\left(\frac{\rho_*(0)}{M_{\odot}~{\rm pc}^{-3}}\right)~{\rm pc}^{-3}~,
\end{equation}
where $f_{CL}$ denotes the fraction of core dark matter in the form of 
molecular clouds. Correspondingly, eq. (\ref{tildeN2}) becomes
\begin{equation}
N_c \simeq  0.13 \times f_{CL}
\left(\displaystyle{\frac{\rho_*(0)}{M_{\odot}~{\rm
pc}^{-3}}}\right)^{1/2}
\left(\frac{M_{DC}}{M_{\odot}}\right)^{1/3}
\left(\frac{r_m}{\rm pc}\right)~,
\end{equation}
on account of eq. (\ref{sigmastella}).
So, the total number of core crossings $N_{cc}$ that a binary 
has to make in order to traverse $N_m$ clouds is
\begin{equation}
N_{cc}\simeq\displaystyle{
\frac{N_m}{N_c}} \simeq 1.4\times 10^{-6}~ f_{CL}^{-1}~
\left(\displaystyle{\frac{M_{\odot}~{\rm
pc}^{-3}}{\rho_*(0)}}\right)^{1/2} 
\left(\frac{\rm pc}{r_m}\right)^2 \left(\frac{t^{(0)}_{21}}{\rm yr}
\right)~.
\end{equation}
Because the core crossing time is 
\begin{equation}
t_{cc}  \simeq 
\left(\frac{9\sigma_*^2}{4\pi G\rho_*(0)}\right)^{1/2}\frac{1}{v}
\simeq  7\times 10^6\left(\frac{M_{\odot}~{\rm pc}^{-3}}{\rho_*(0)}
\right)^{1/2}~{\rm yr}~,
\end{equation}
it follows that the process under consideration takes a total time
$N_{cc} t_{cc}$. However, this time is by definition just $t_{21}$.
So, we get
\begin{equation}
t_{21}  \simeq  
N_{cc}~t_{cc} 
\simeq  9.8~
f^{-1}_{CL}
\left(\frac{M_{\odot}~{\rm pc}^{-3}}{\rho_*(0)} \right ) 
\left( \frac{\rm pc}{r_m}\right)^2 t^{(0)}_{21}~,
\label{fcl}
\end{equation}
which is the desired relationship between $t_{21}$ and $t^{(0)}_{21}$.
We are now in position to re-express eq. (\ref{a6}) in terms of $t_{21}$.
Accordingly, we get
\begin{equation}
\left(\frac{{\rm km}}{a_2}\right)^{3/2} \simeq   
\left(\frac{{\rm km}}{a_1}\right)^{3/2} +2.1\times 10^{-27} f_{CL}~
\Xi^{-1}
\left(\frac{\rho_*(0)}{M_{\odot}~{\rm pc}^{-3}}\right )
\left(\frac{t_{21}}{\rm yr}\right)~.
\label{a7}
\end{equation}

In order to quantify the effect of frictional hardening, we may proceed 
much in the same way as in the case of collisional hardening, neglecting 
however the latter effect for the moment. Specifically, taking $t_{21}$
in eq. (\ref{a7}) equal to the age of the Universe, we find that the
present orbital radius of a binary brown dwarf is given by
\begin{equation}
\left(\frac{{\rm km}}{a_2}\right)^{3/2} \simeq   
\left(\frac{{\rm km}}{a_1}\right)^{3/2} +2.1\times 10^{-17}f_{CL}~
\Xi^{-1}
\left(\frac{\rho_*(0)}{M_{\odot}~{\rm pc}^{-3}}\right )~.
\label{31}
\end{equation}
Manifestly, frictional hardening is operative to the extent that $a_2$
becomes considerably smaller than $a_1$. Accordingly, from eq. (\ref{31}) we
see that this is indeed the case provided
\begin{equation}
a_1  \ut  > 1.3\times 10^{11} f_{CL}^{-2/3}~\Xi^{2/3}
\left(\frac{M_{\odot}~{\rm pc}^{-3}}{\rho_*(0)}\right)^{2/3}
~~{\rm km}~.
\label{100}
\end{equation}
Owing to eq. (\ref{100}), eq. (\ref{31}) entails in turn
\begin{equation}
a_2 \simeq  1.3\times 10^{11} f_{CL}^{-2/3}~\Xi^{2/3}
\left(\frac{M_{\odot}~{\rm pc}^{-3}}{\rho_*(0)}\right)^{2/3}~~{\rm km}~.
\label{33}
\end{equation}
Physically, only those 
binaries whose initial orbital radius satisfies condition (\ref{100}) are
affected by frictional hardening, and their present orbital radius
turns out to be almost {\it independent} of the initial
value. We can make the present discussion somewhat more specific by
noticing that our assumptions strongly suggest
$\rho_*(0)\ut < 10^4~M_{\odot}~{\rm pc}^{-3}$, in which case
both eq. (\ref{100}) and eq. (\ref{33}) acquire the form
\begin{equation}
a_{1,2}\ut > 2.8 \times 10^{8} f_{CL}^{-2/3}~\Xi^{2/3}~~~{\rm km}~.
\label{34}
\end{equation}
Evidently, very hard primordial
binaries -- which violate condition (\ref{34}) -- do not suffer frictional
hardening, and the same is true for tidally-captured binaries. 

\subsubsection{Present orbital radius of primordial binaries}

Which of the two hardening mechanisms under consideration is more
effective? A straightforward implication of eqs. (\ref{a4}) 
and (\ref{33})
is that frictional hardening is more efficient than collisional
hardening whenever it so happens that
\begin{equation}
f_{IBD} < 6.3 \times 10^{-2} f^{2/3}_{CL}~ \Xi^{-2/3} 
\left(\frac{M_{DC}}{M_{\odot}}\right)^{1/3} 
\left(\frac{M_{\odot} {\rm pc}^{-3}}{\rho_{*}(0)}\right)^{1/3}~.
\label{74}
\end{equation}
Unfortunately, the occurrence of various dark cluster parameters in 
condition (\ref{74}) does not permit a sharp conclusion to be drawn, but 
in the illustrative case $M_{DC} \simeq 10^5 M_{\odot},
r_m \simeq 10^{-3}~ {\rm pc}, f_{CL} \simeq 0.5$ and 
$\rho_*(0) \simeq 10^3 M_{\odot}~ {\rm pc}^{-3}$ we get $f_{IBD} < 0.5$.
As we expect in the cores $f_{IBD} \ll f_{PB}$ and 
$f_{CL} \simeq f_{PB}$, we see that {\it frictional hardening
plays the dominant r\^ole}. Moreover, from eqs. (\ref{a4}) and (\ref{33})
it also follows that the effectiveness of collisional hardening 
decreases for smaller values of $f_{IBD}$. Finally, the fairly slow 
dependence of condition (\ref{74}) on $M_{DC}$ and $\rho_*(0)$ makes
our conclusion rather robust.

Having shown that collisional hardening can effectively be disregarded,
we get that the present orbital radius of primordial binary brown dwarfs is
actually given by eq. (\ref{33}), as long as condition
(\ref{100}) is met. Taking again the above particular
case as an illustration, we find $a_2 \simeq 7 \times 10^8$ km, which
is of the same order as the Einstein radius for microlensing towards
the LMC (Gaudi and Gould \cite{gg}). Therefore, we argue that
not too hard primordial binaries can be resolved in future microlensing 
experiments with a more accurate photometric observation, the
signature being small deviations from standard microlensing
light curves (Dominik \cite{dominik}).

\subsubsection{Gravitational encounters}

We are now in position to take up the question concerning the survival
of binary brown dwarfs against gravitational encounters. As already
pointed out, individual-binary encounters are harmless in this 
respect -- since hard binaries are considered throughout -- and so
we shall restrict our attention to binary-binary encounters.

As a first step, we recall that their average
rate $\Gamma$ 
(Spitzer \& Mathieu 1980) can presently be written as
\begin{equation}
\Gamma ~ \simeq~ \alpha~ \frac{G m a}{\sigma_*}~,
\label{7}
\end{equation}
with $\alpha \simeq 13$. Correspondingly, on account 
of eq. (\ref{sigmastella})
the reaction time in the dark cluster cores turns out to be
\begin{equation}
t_{\rm react} \simeq 10^{19} \beta \left(\frac{M_{\odot}~{\rm pc}^{-3}}
{\rho_*(0)} \right) \left(\frac{M_{DC}}{M_{\odot}} \right)^{1/3} 
\left( \frac{{\rm km}}{a} \right) ~{\rm yr}~,
\label{60}
\end{equation}
with $\beta \simeq 6.7/f_{PB}$. 

Now, if no hardening were to occur -- which means that the orbital 
radius $a$ would stay constant -- the binary survival condition would
simply  follow by demanding that $t_{react}$ should exceed the age
of the Universe.
However, hardening makes $a$ decrease, and so $t_{react}$ increases
with time. This effect can be taken into account by considering the
averge value $< t_{react} >$ of the reaction time over the time
interval in question (to be denoted by T), namely 
\begin{equation}
< t_{react} >~ \equiv ~\frac{1}{T} ~ \int_0^T dt~t_{react}~.
\label{aver}
\end{equation}
In order to compute $< t_{react} >$, the temporal dependence
$a(t)$ of the binary orbital radius is needed. 
Because frictional hardening plays the major r\^ole, the latter
quantity is evidently supplied by eq. (\ref{a7}).
Setting for notational convenience $t \equiv t_{21}$
and $a(t) \equiv a_2$, eq. (\ref{a7}) yields 
\begin{equation} 
a(t) \simeq  \left[ \left(\frac{{\rm km}}{a_1}\right)^{3/2}
+ 2.1 \times 10^{-27}~ \Xi^{-1} f_{CL}
\left(\frac{\rho_*(0)}{M_{\odot}
{\rm pc^{-3}}} \right)\left(\frac{t}{\rm yr}\right ) \right]^{-2/3}
~{\rm km}~.
\label{n1}
\end{equation}
Combining eqs. (\ref{60}) and (\ref{n1}) together and inserting the
ensuing expression into eq. (\ref{aver}), we get
\begin{eqnarray}
< t_{react} > & \simeq & 1.9 \times 10^{46}~ f^{-1}_{PB}
f^{-1}_{CL}~ \Xi \left( \frac{M_{DC}}{M_{\odot}}\right)^{1/3} 
\left(\frac{{\rm yr}}{T}\right) 
\left(\frac{{\rm km}}{a_1} \right)^{5/2}
\left(\frac{M_{\odot} {\rm pc^{-3}}}{\rho_*(0)}\right)^2
\times \nonumber \\ & &
\left\lbrace \left[1+2.1\times 10^{-27} \Xi^{-1} f_{CL} 
\left(\frac{\rho_*(0)}{M_{\odot} {\rm pc^{-3}}}\right)
\left(\frac{T}{\rm yr}\right) \left(\frac{a_1}{\rm km}\right)^{3/2}
\right]^{5/3} -1 \right\rbrace~{\rm yr}.
\label{n2}
\end{eqnarray} 
Let us now require $< t_{react} >$ to exceed the age of the
Universe (taking evidently $T \simeq 10^{10}$ yr). As it is apparent
from eq. (\ref{60}), $t_{react}$ is shorter for softer binaries. 
Hence, in order to contemplate {\it hard} binaries with an arbitrary orbital
radius we have to set $a_1 \simeq 1.4 \times 10^{12} (M_{\odot}/M_{DC})^{2/3}$
km in eq. (\ref{n2}). Correspondingly, the binary survival condition
reads \footnote{Application of the same argument to tidally - captured
binaries shows that no depletion occurs in this way.}
\begin{eqnarray}
f_{PB} & \ut <  & 8.2 \times 10^{-5} f_{CL}^{-1}~ \Xi
\left(\frac{M_{DC}}{M_{\odot}}\right)^2 \left( \frac{M_{\odot}{\rm pc^{-3}}}
{\rho_*(0)} \right)^2 \times \nonumber \\ & &
\left\lbrace
\left[1+ 35 f_{CL} \Xi^{-1} \left(\frac{\rho_*(0)}{M_{\odot}{\rm pc^{-3}}}
\right) \left(\frac{M_{\odot}}{M_{DC}}\right)\right]^{5/3} -1
\right\rbrace~.
\label{n3}
\end{eqnarray}
Although the presence of various dark cluster parameters prevents a 
clear - cut conclusion to be drawn from eq. (\ref{n3}), in the
illustrative case $M_{DC} \simeq 10^5 M_{\odot}$ and $f_{CL} \simeq 0.5$
eq. (\ref{n3}) entails e.g. $f_{PB} \ut < 0.3$ for
$\rho_*(0) \simeq 3 \times 10^3 M_{\odot}~ {\rm pc^{-3}}$. Thus, we infer
that for realistic values of the parameters in question a sizable fraction of 
primordial binary brown dwarfs survives binary-binary encounters in the
dark cluster cores \footnote{If the initial value of $f_{PB}$ fails to satisfy 
condition (\ref{n3}), primordial binaries start to be destroyed in 
binary-binary encounters until 
their fractional abundance gets reduced down to a value
consistent with condition (\ref{n3}).}.

\section{Thermal balance in halo molecular clouds}
As far as the energetics of halo molecular clouds is concerned,
we can identify two main heat sources. 
One is energy deposition from background 
photons, while the other is of gravitational origin (coming from frictional
hardening of primordial binary brown dwarfs). Although we shall discuss
them separately, it should be kept in mind that they both act at the same
time.

\subsection{Energy from background photons}
We proceed to estimate $T_m$
by momentarily neglecting gravitational effects.
To this end, we need to know the heating rate
(due to external sources) and the cooling rate (due to the 
molecules). In the galactic halo, the dominant heat source for molecular 
clouds is expected to be ionization from photons
of the X-ray background, whose spectrum in the relevant 
range $1~ {\rm keV} < E < 25$ keV (see below)
can be parametrized in terms of the 
energy $E$ (expressed in keV) as 
$I(E)= 8.5~ E^{-0.4}~~{\rm cm^{-2}~s^{-1}~sr^{-1}}$ 
(O'Dea et al. \cite{ODea}).
The ionization rate per $H_2$ molecule (taking secondary ionization
into account) is 
\begin{equation}
\xi_X = 26~\int_{E_{min}}^{E_{max}} 4 \pi I(E) \sigma_X(E) dE~~{\rm s^{-1}},
\label{eqno:xi}
\end{equation}
where  $\sigma_X(E) = 2.6 \times 10^{-22}~E^{-8/3}$ cm$^2$ 
is the absorption cross-section in the above energy range
for incoming X-rays on gas with interstellar 
composition (Morrison and McCammon \cite{Morrison}).
Actually,  $\sigma_X(E)$ decreases faster as $E$ increases 
when the gas metallicity is lower (see Fig.1 in Morrison and 
McCammon \cite{Morrison}).  So, the above quoted expression  for
$\sigma_X(E)$ is expected to yield un upper bound on the cross-section
for X-rays on halo molecular clouds. 
The integration limits $E_{min}$ (below which
X-rays are absorbed) and $E_{max}$ (above which X-rays 
go through the whole cloud without being absorbed)
depend on the cloud column density.
Thus, taking as an orientation $n_m$ in the range  
$10^4 - 10^8$~ cm$^{-3}$,
we get $1.25~ {\rm keV} < E_{min} < 7$ keV and
$10~{\rm keV} < E_{max} < 20~{\rm keV}$.
It turns out that $\xi_X$ is rather insensitive to the upper limit, and
we obtain 
$\xi_X = 2.2 \times 10^{-19}~{\rm s^{-1}}$ 
for $E_{min}=1.25$ keV and
$\xi_X = 5.6 \times 10^{-21}~{\rm s^{-1}}$
for $E_{min}=7$ keV
\footnote{
It is straightforward to verify that the ionization rate $\xi_{cr}$ 
due to the halo cosmic rays is less important in this context. 
In fact, assuming the ionization rate per $H_2$ molecule
typical for disk 
vcosmic rays $\xi_0 = 5 \times 10^{-17}$ s$^{-1}$ (see e.g. van Dishoeck and
Black \cite{vand}) and 
rescaling for the diffusion of the cosmic rays in the galactic halo 
(De Paolis et al. \cite{depaolis2}), 
we infer $\xi_{cr} \sim 10^{-23} ~{\rm s^{-1}}$.}.
Since each ionization process releases an energy of $\simeq 8~{\rm eV}$, 
the heating rate per $H_2$ molecule 
$\Gamma$ turns out to be
\begin{equation}
3.5\times 10^{-32}~{\rm erg}~{\rm s}^{-1}/H_2~
 < \Gamma < 1.5\times 10^{-30}~{\rm erg}~ {\rm s}^{-1}/H_2~.
\end{equation}

Unfortunately, the 
cooling rate for the halo clouds in question is not well known,
owing to the lack of detailed information on their 
chemical composition.
Nevertheless, by merely considering the cooling rate due to $H_2$ 
as given by Goldsmith and Langer \cite{Goldsmith}, 
the equality between cooling and 
heating rate per molecule leads to $T_m \simeq 10$ K.
More complete models for the cooling rate,
which include the contribution from $HD$ and heavy molecules, 
imply that the cooling efficiency is substantially enhanced 
and thus make it very plausible that halo 
molecular clouds clumped into dark clusters should have a 
temperature close to that of the CBR, namely $T_m \simeq 3$ K
(see also Gerhard and Silk \cite{Gerhard} for similar
conclusions).

\subsection{Energy from primordial binary brown dwarfs}
As the analysis in Sect 4.3 shows, dynamical friction 
transfers a huge amount of 
energy from primordial binary brown dwarfs to molecular clouds, and so
it looks compelling to investigate (at least) the gross features of
the ensuing energy balance.

Let us start by evaluating the energy  acquired by molecular
clouds in the process of frictional hardening. Recalling that the traversal
time for a single cloud is given by eq. (\ref{16}), eq. (\ref{a6}) 
entails that --
after a binary with initial orbital radius $a_1$ 
has crossed $N$ clouds -- its orbital radius gets reduced down
to
\begin{equation}
%\begin{array}{l}
{a_{N+1}} \simeq \left[ 
1.1\times 10^{-19}N~\Xi^{-1}\left(\frac{\rm pc}{r_m}\right)
\left(\frac{M_{\odot}}{M_{DC}}\right)^{1/3} + 
\left(\frac{{\rm km}}{a_1}\right)^{3/2} \right]^{-2/3}~{\rm km}~. 
\label{29}
%\end{array}
\end{equation}
Accordingly, we see that the orbital radius remains almost constant
until $N$ reaches the critical value
\begin{equation}
N_0\equiv 9\times 10^{18}~\Xi
\left(\frac{r_m}{{\rm pc}} \right)
\left(\frac{M_{DC}}{M_{\odot}} \right)^{1/3}
\left(\frac{{\rm km}}{a_1}\right)^{3/2}~,
\end{equation}
whereas it {\it decreases} 
afterwards. Because the
energy acquired by the clouds is just the binding energy given up by 
primordial binaries, this information can be directly used to
compute the energy $\Delta E_c(N)$ gained by the $N$-th cloud traversed
by a binary whose initial orbital radius was $a_1$. 
Manifestly, we have
\begin{equation}
\Delta E_c(N)=\frac{1}{2}Gm^2
\left(\frac{1}{a_{N+1}}-\frac{1}{a_N} \right)~.
\end{equation}
Thanks to eq. (\ref{29}), 
a straightforward calculation shows that $\Delta E_c(N)$
stays practically constant
\begin{equation}
\Delta E_c \simeq 9.8\times 10^{32}~\Xi^{-1}
\left(\frac{{\rm pc}}{r_m} \right)
\left(\frac{M_{\odot}}{M_{DC}} \right)^{1/3}
\left(\frac{a_1}{{\rm km}}\right)^{1/2}~~{\rm erg}~,
\label{32}
\end{equation}
as long as $N\ut < N_0$, while it subsequently {\it decreases}. 
So, the
amount of energy transferred to a cloud is maximal during the early stages 
of hardening. Now, since the binding energy of a cloud is
\begin{equation}
E_c\simeq 7.7\times 10^{42}
\left(\frac{r_m}{{\rm pc}} \right)~~{\rm erg}~,
\end{equation}
it can well happen that $\Delta E_c >E_c$ (depending on $r_m$, $M_{DC}$ 
and $a_1$), which means that the cloud would evaporate unless it
manages to efficiently dispose of the excess energy.

A deeper insight into this issue can be gained as follows
(we focus the attention on the early stages of hardening, when the
effect under consideration is more dramatic).
Imagine that a spherical cloud is crossed by a primordial 
binary which moves along a straight line, and consider the cylinder $\Delta$
traced by the binary inside the cloud (its volume being approximately
$\pi a^2 r_m$). 
Hence, by $n_m \simeq 62.2 ~({\rm pc/r_m})^2 ~
{\rm cm}^{-2}$ (which follows from eq. (12)) the average
number of molecules inside $\Delta$ turns out to be
\begin{equation}
N_{\Delta} \simeq 5.8 \times 10^{30} 
\left( \frac{a_1}{{\rm km }}\right)^2
\left( \frac{{\rm pc}}{ r_m}\right)~.
\label{NDelta}
\end{equation}
Physically, the energy $\Delta E_c$ gets first deposited within $\Delta$
in the form of heat. Neglecting thermal conductivity (more about this,
later),  the temperature 
inside $\Delta$ accordingly becomes
\begin{equation}
T_{\Delta} \simeq  \frac{2}{3}\frac{\Delta E_c}{N_{\Delta}k_B} 
\simeq 8.1 \times 10^{17}~\Xi^{-1} 
\left( \frac{M_{\odot}}{M_{DC}}\right)^{1/3}
\left( \frac{{\rm km }}{a_1}\right)^{3/2}~~{\rm K}~,
\label{TDelta}
\end{equation}
$k_B$ being the Boltzmann constant.
On account of eqs. (\ref{a1}) and (\ref{33}), eq. (\ref{TDelta}) yields
\begin{equation}
0.5~\Xi^{-1}\left(\displaystyle{\frac{M_{DC}}{M_{\odot}}}\right)^{2/3}~{\rm K}
~\ut < ~T_{\Delta} ~\ut < 
16.2~ 
f_{CL}~\Xi^{-2}  \left(\displaystyle{\frac{\rho_*(0)}{M_{\odot}{\rm pc^{-3}}}}
\right)
\left(\displaystyle{\frac{M_{\odot}}{M_{DC}}}\right)^{1/3}~{\rm K}~,
\label{ccc}
\end{equation}
which -- in the illustrative case $f_{CL}\simeq 0.5$, 
$\rho_*(0) \simeq 3 \times 10^3 M_{\odot}~ {\rm pc^{-3}}$
and $M_{DC}\simeq 10^5~M_{\odot}$ -- entails in turn $5.3\times 10^3~
{\rm K}~\ut < T_{\Delta}\ut < ~1.3\times 10^4~{\rm K}$.
As a consequence of the increased temperature, the molecules within 
$\Delta$ will radiate, thereby reducing the excess energy in the
cloud. In order to see whether this mechanism actually prevents the
cloud from evaporating, we notice that the characteristic time needed
to accumulate the energy $\Delta E_c$ inside $\Delta$ is just the
traversal time $t_m$. Therefore, this energy will be totally radiated
away provided the cooling rate per molecule $\Lambda$ exceeds the critical
value $\Lambda_0$ given by the equilibrium condition
\begin{equation}
N_{\Delta} \Lambda_0 t_m \simeq \Delta E_c~.  \label{37}
\end{equation}
Specifically, eq. (\ref{37}) yields
\begin{equation}
\Lambda_0 \simeq 10^{-12}~ \Xi^{-1} \left(\frac{{\rm pc}}{r_m} \right)
\left(\frac{{\rm km}}{a_1}\right)^{3/2} ~{\rm erg~ s^{-1}~ mol^{-1}}~,
\label{338}
\end{equation}
on account of eqs. (\ref{16}), (\ref{32}) and 
(\ref{NDelta}). Moreover, in the present case
in which most of the molecules are $H_2$ the explicit form of
$\Lambda$ is 
(see e.g. O'Dea et al. 1994, Neufeld et al. 1995)
\begin{equation}
\Lambda \simeq 3.8 \times 10^{-31}~\left(\frac{T_{\Delta}}
{{\rm K}}\right)^{2.9}
~~{\rm erg~s^{-1}~H_2^{-1}}~, 
\label{27}
\end{equation}
which - thanks to eq. (\ref{TDelta}) - becomes
\begin{equation}
\Lambda \simeq 3.4 \times 10^{21} ~\Xi^{-2.9} \left(\frac{{\rm km}}
{a_1} \right)^{4.35}~ {\rm erg~s^{-1}~H_2^{-1}} \label{40}
\end{equation}
(notice that $\Lambda$ is almost independent of $M_{DC}$). Now,
from eqs. (\ref{338}) and (\ref{40}) it follows that the condition
$\Lambda \ut > \Lambda_0$ implies
\begin{equation}
a_1 \ut < ~6 \times 10^{11} ~\Xi^{-0.7} 
\left(\frac{r_m}{{\rm pc}}\right)^{0.35} ~{\rm km}~.
\label{410}
\end{equation}

%Again, it is difficult to figure out the relevance of condition (\ref{410})
%in general, but in the illustrative case of $M_{DC} \simeq 10^5
%M_{\odot}$ it turns out to be abundantly met for hard primordial binaries.
For a wide range of dark cluster and molecular cloud parameters it
turns out that eq. (\ref{410}) is fulfilled for hard primordial binaries.

Thus, we conclude that
the energy given up by primordial binary brown dwarfs and 
temporarily acquired by molecular clouds is efficiently radiated away,
so that the clouds do not get dissolved by frictional hardening.

As a final comment, we stress that our estimate for $T_{\Delta}$
should be understood as an upper bound, since thermal conductivity
has been neglected. In addition,  
the above analysis implicitly
relies upon the assumption $T_{\Delta} < 10^4$ K, which ensures the
survival of $H_2$. Actually, in spite of the fact that condition (\ref{ccc})
entails that this may well not be the case, our conclusion remains  
nevertheless true. For, higher temperatures would lead to the depletion
of $H_2$, which correspondingly gets replaced by atomic and
possibly ionized hydrogen. As is well known, in either case the resulting
cooling rate would exceed the one for $H_2$, and so cooling would be
even more efficient than estimated above (B\"ohringer \& Hensler 1989).

\section{Lyman-$\alpha$ absorption systems}
It is well known that Quasar Ly-$\alpha$ absorption lines provide a
detailed information on the evolution of the gaseous component of 
galaxies (see e.g. Fukugita, Hogan and Peebles \cite{fhp} and 
references therein). These lines are seen for a neutral hydrogen 
column density $N_{HI}$
ranging from $\sim 3 \times 10^{12}$ cm$^{-2}$ (the detection threshold) to 
$\sim 10^{22}$ cm$^{-2}$. 

At the upper limit of this range ($N_{HI} \ge 2 \times 10^{20}$ cm$^{-2}$) 
the lines are classified as damped Ly-$\alpha$ systems (mostly associated with 
metal-rich objects) and it is generally believed that they are the thick 
progenitors of galactic disks. The $HI$ distribution in damped Ly-$\alpha$ 
systems is usually flatter 
than the corresponding surface brightness of the optical disks and 
extends to much larger radii (up to $\sim 40$ kpc in giant galaxies). 
In the outer galactic regions Ly-$\alpha$ systems show sharp $HI$ edges at a 
level of $N_{HI} \sim 10^{17}$ cm$^{-2}$, the so-called Lyman limit.
Damped Ly-$\alpha$ systems are observed up to redshift $z \sim 3.5$ and 
the survey results suggest that the average mass of neutral hydrogen per 
absorption system decreases with time, in agreement with the hypothesis of 
gas consumption into stars (Lanzetta, Wolfe and Turnshek \cite{lwt}). 

Within our picture, it is tempting to identify
damped Ly-$\alpha$ systems  
with the PGC clouds in the inner galactic halo, where they undergo
disruption.

Below the Lyman limit (i.e. for $3 \times 10^{12}~ {\rm cm}^{-2} < 
N_{HI} < 3 \times 10^{15}$~ cm$^{-2}$, see e.g. Fig. 2 in 
Cristiani \cite{cristiani}), 
the absorption lines are 
classified as Ly-$\alpha$ forest and are generally ascribed to a 
large number of intervening clouds (in some cases extending up to $\sim 200$ 
kpc from a central galaxy) along the line of sight to distant Quasars.
Studies of Ly-$\alpha$ forest lines have made rapid progress recently and 
some observational trends are today firmly established. In particular:
$(i)$ The evolution of the co-moving number density of systems 
per unit interval in redshift shows that the number of Ly-$\alpha$ 
forest clouds rapidly decreases with time for $z>2$, while it is approximately 
constant for $z<2$.
$(ii)$ The correlation between the thermal Doppler parameter $b$ and the 
number of clouds can be fitted with a gaussian distribution of median 
$<b>~\simeq 30$ km s$^{-1}$ and dispersion $\simeq 10$ km s$^{-1}$, 
corresponding to a temperature of a few $10^4$ K.

Within our model, it looks natural to identify Ly-$\alpha$ forest 
clouds with the molecular clouds clumped into dark clusters located 
in the outer halo.
Indeed, we expect that the clouds contain
in their external layers an increasing fraction of $HI$ gas and that the outer
regions are even ionized, due to the incoming UV radiation
\footnote{Observe that so far we have been assuming that halo molecular
clouds have a constant temperature. This was a good approximation as
far as the previous analysis was concerned, but would be too poor
in the present discussion}.
Observe that so far we have been assuming that halo molecular clouds
have a constant temperature. This was a good approximation as far
as the previous analysis was concerned, but would be too poor in the
present discussion.
A $HI$ column density
of $\sim 10^{14}$ cm$^{-2}$ -- corresponding to a layer of about
$10^{-6}$ pc -- is sufficient to shield the incoming radiation. 
Remarkably enough, $N_{HI} \sim 10^{14}~ {\rm cm}^{-2}$
corresponds to the average value of the observed Lyman-$\alpha$ forest
distribution (Cristiani \cite{cristiani}).
Since we expect the
UV flux to decrease as the galactocentric distance increases,
clouds which lie at a larger distance may thus have a smaller
$HI$ column density, whereas clouds closer to the galactic disk
may have a higher column density. This fact explains
the observed distribution for the column density.
%A value of $< b > \sim 30$ km/s, corresponding to a temperature of
%$\sim 10^4$ K, is the expected temperature of the atomic layer
%of our halo clouds.   
Moreover, the number of clouds is expected to decrease
with time initially, due to both MACHO and $H_2$  formation, in
this way explaining the observed evolution of Ly-$\alpha$ forest clouds
according to the above-mentioned point $(i)$.
As a final comment, we mention that very recently R\"ottgering, Miley and 
Van Ojik 
\cite{rmv}, by considering the filling factors and the physical parameters 
derived from their Ly-$\alpha$ forest observations, pointed out that 
galactic halos may contain $\sim 10^9~M_{\odot}$ of 
neutral hydrogen gas and are 
typically composed of $\sim 10^{12}$ clouds, each of size $\sim$ 
40 light-days. It looks remarkable that
these are the typical parameters of molecular 
clouds dealt with in this paper.

A thorough quantitative analysis requires, however, further
investigations, which are beyond the scope of the present paper.

\section{Conclusions}
Looking back at what we have done, it seems to us fair to say that present-day
results of microlensing experiments towards the LMC 
are amenable of a very simple explanation
in terms of what was repeatedly proposed as a natural candidate
for baryonic dark matter, namely brown dwarfs.
Once this idea is accepted, a few almost obvious steps just follow.
First, given the fact that ordinary halo stars form in
(globular) clusters, it seems more likely that brown dwarfs
as well  form  in  clusters rather than in isolation
(this circumstance has been repeatedly recognized in the last few years). 
Of course, 
whether brown dwarfs are still clumped into dark clusters today is
a different and nontrivial question -- we have seen that core collapse can
liberate a considerable fraction of brown dwarfs from the less massive
dark clusters. Second, much in the same way as it happens for ordinary
stars, a consistent fraction of binary brown dwarfs (up to $50\%$ in mass) 
is expected to form.
Third, since no known star-formation mechanism is very
efficient, it is natural to imagine that a substantial amount
of primordial gas is left over. Although we cannot be sure about the
subsequent fate of this gas -- which should be mostly cold $H_2$ --
it is likely to remain confined within the dark clusters, since
stellar winds are absent.
So, also cold $H_2$ self-gravitating clouds should presumably be
clumped into dark clusters, along with some residual diffuse gas.

We have shown that -- within the considered model -- 
not too hard primordial binary brown dwarfs turn out to have an
orbital radius which is typically of the order of the Einstein radius
for microlensing towards the LMC. Therefore, they are not so easily
resolvable in the microlensing experiments performed so far, but
we argue that they can be resolved in
future microlensing experiments with  a more accurate photometric
observation
\footnote{
See e.g. the ongoing experiments by the GMAN and PLANET collaborations
(Proceedings of the {\it Second International Workshop on Gravitational 
Microlensing Surveys}, Orsay, 1996).}, the signature being small
deviations from standard microlensing light curves (Dominik \cite{dominik}).
Notice that such a procedure complements the detection strategy for
binaries suggested by Gaudi and Gould \cite{gg}. 

Many physical processes certainly occur in the dark clusters, and the
(likely) presence of a large amount of gas makes them even more
difficult to understand than globular clusters. Moreover, the 
lack of any observational information makes any attempt to figure out
the physics of dark clusters almost impossible, were it not for some
remarkable structural analogies with globular clusters that the model
-- upon which our discussion is based --
naturally suggests. What should anyway be clear is that a host of
different phenomena can happen. Some of them have intentionally
been neglected here, in order not to make the present paper either
excessively long or too speculative. Yet, many others are
likely to occur, that we have not even been able to imagine.

\section{Acknowledgements}
We would like to thank B. Bertotti, B. Carr 
and F. D'Antona for useful discussions. FDP has been partially supported
by the Dr. Tomalla Foundation and by INFN. 
FDP and GI acknowledge some support from ASI
(Agenzia Spaziale Italiana).


\begin{thebibliography}{}
\bibitem[1996]{adams}
Adams, F.C. \& Laughlin, G. 1996, ApJ 468, 586
\bibitem[1993]{alcock}
Alcock, C. et al. 1993, Nat 365, 621
\bibitem[1997]{Alcock1}
Alcock, C. et al. 1997, ApJ 486, 697
\bibitem[1990]{ashman} 
Ashman,K. M. 1990, MNRAS 247, 662
\bibitem[1962]{antonov}
Antonov, V. A. 1962, Vestn Leningrad-Gross. Univ. 7, 135
\bibitem[1993]{aubourg}
Aubourg, E. et al. 1993, Nat 365, 623
\bibitem[1994]{bahcall}
Bahcall, J., Flynn, C., Gould, A., \& Kirhakos, S. 1994, ApJ 435,
L51
\bibitem[1984]{bs}
Bettwieser, E. \& Sugimoto, M. 1984, MNRAS 208, 439
\bibitem[1987]{binney}
Binney, J. \& Tremaine, S. 1987 {\it Galactic Dynamics} Princeton
University Press, Princeton
\bibitem[1989]{bor}
B\"ohringer, H. \& Hensler, G. 1989, A\&A 215, 147
\bibitem[1989]{burrows} 
Burrows, A., Hubbard, W. B. \& Lunine, J. I. 1989, ApJ 345, 939
\bibitem[1994]{bcarr}
Carr, B. 1994, Ann. Rev. Astron. Astrophys. 32, 531
\bibitem[1984]{RCarr} 
Carr, B. J., Bond, J. \& Arnett, W. 1984, ApJ  277, 445 
\bibitem[1987]{lacey} 
Carr, B. J. \& Lacey, C. G. 1987, ApJ 316, 23
\bibitem[1996]{chabrier}
Chabrier, G., Segretain, L. \& M\'era, D 1996, ApJ 468, L21
\bibitem[1995]{charly}
Charlot, S. \& Silk, J. 1995, ApJ 445, 124
\bibitem[1980]{ch}
Cohn, H. 1980, ApJ 242, 765
\bibitem[1984]{chh}
Cohn, H. \& Hut, P. 1984, ApJ 277, L45
\bibitem[1996]{cristiani}
Cristiani, S. 1996, ESO preprint 1117 
\bibitem[1972]{Dalgarno} 
Dalgarno, A \&  McCray, R. A. 1972, ARAA 10, 375
\bibitem[1987]{D'Antona} 
D'Antona, F. 1987, ApJ 320, 653
\bibitem[1995a]{depaolis1}
De Paolis, F., Ingrosso, G., Jetzer, Ph. \& Roncadelli, M. 1995a, Phys. Rev. 
Lett. 74, 14 
\bibitem[1995b]{depaolis2}
De Paolis, F., Ingrosso, G., Jetzer, Ph. \& Roncadelli, M. 1995b, A\&A 
295, 567
\bibitem[1995c]{depaolis3}
De Paolis, F., Ingrosso, G., Jetzer, Ph., Qadir, A. \&
Roncadelli, M. 1995c, A\&A 299, 647
\bibitem[1995d]{depaolis4}
De Paolis, F., Ingrosso, G., Jetzer, Ph. \& Roncadelli, M. 1995d,
Comments on Astrophys. 18, 87
\bibitem[1996]{Ingrosso} De Paolis, F., Ingrosso, G. \& Jetzer, Ph. 1996,
ApJ 470, 493
\bibitem[1991]{Derujula} De R\'ujula, A., Jetzer, Ph. \& Mass\'o, E. 1991,
MNRAS 250, 348
\bibitem[1992]{der}
De R\'ujula, A., Jetzer, Ph. \& Mass\`o, E. 1992, A\&A 254, 99
\bibitem[1990]{DickeyLockman}
Dickey, J. M. \& Lockman,  F. J. 1990, ARAA, 28, 215
\bibitem[1996]{dominik}
Dominik, M. 1996, Thesis, Dortmund University
\bibitem[1991]{Einaudi}  
Einaudi, G. \& Ferrara, A. 1991, ApJ 371, 571 
%\bibitem[1989]{Elmegreen}
%Elmegreen, B. G. 1989, ApJ, 338, 178
\bibitem[1996]{evans}
Evans, N.W. 1996, to appear in 
{\it Aspects of dark matter in astro and particle physics},
ed. by Klapdor-Kleingrothaus, H. V. (World Scientific, Singapore)
\bibitem[1975]{fabian}
Fabian, A. C., Pringle, J. E. \& Rees, M. J. 1975, MNRAS 172, 15
\bibitem[1994]{fn}
Fabian, A.C. \& Nulsen, P.E.J. 1994, MNRAS 269, L33
\bibitem[1985]{Fall}
Fall, S. M. \& Rees, M. J. 1985, ApJ 298, 18
\bibitem[1996]{fields}
Fields, B. D., Mathews, G. \& Schramm, D. N. 1996, astro-ph 9603035
\bibitem[1996]{fhp}
Fukugita, M., Hogan, C. J. \& Peebles, P. J. E. 1996, Nat 381, 489
\bibitem[1996]{gates}
Gates, E.J., Gyuk, G. \& Turner, M. 1996, Phys. Rev. D53, 4138
\bibitem[1997]{gg}
Gaudi, B.S. \& Gould, A. 1997, ApJ 482, 83
\bibitem[1996]{Gerhard}
Gerhard, O. E. \& Silk, J. 1996, ApJ 472, 34
\bibitem[1997]{gm}
Gibson, B.K. \& Mould, J.R. 1997, ApJ 482, 98
\bibitem[1978]{Goldsmith}
Goldsmith P. F. \& Langer, W. D. 1978, ApJ 222, 881
\bibitem[1996]{graff}
Graff, D. S. \& Freese, K. 1996, ApJ 456, L49
\bibitem[1975]{heggie}
Heggie, D. C. 1975, MNRAS 173, 729
\bibitem[1989]{hr}
Heggie, D. C. \& Ramamani, N. 1989, MNRAS 237, 757
\bibitem[1969]{henon}
H\'enon, M. 1969, A\&A 2, 151
\bibitem[1994]{hu}
Hu, E. M., Huang, J. S., Gilmore, G. \& Cowie, L. L. 1994, Nat 371, 493
\bibitem[1996]{Jetzer} 
Jetzer, Ph. 1996, Helv. Phys. Acta 69, 179
\bibitem[1990]{Kang} 
Kang, H., Shapiro, P. R., Fall, S. M. \& Rees, M. J. 1990, ApJ 363, 488
\bibitem[1996]{kawaler}
Kawaler, S. D. 1996, ApJ 467, L61
\bibitem[1996]{kerins}
Kerins, E. J. 1996, astro-ph 9610070
\bibitem[1997]{kerins97}
Kerins, E. J. 1997, astro-ph 9704179
\bibitem[1995]{lwt}
Lanzetta, K. M., Wolfe, A. M. \& Turnshek, D. A. 1995, ApJ 440, 435
\bibitem[1990]{larson}
Larson, R.B. 1990, in {\it Dynamics and Interactions of Galaxies}
ed. Wielen, R. (Springer, Berlin)
\bibitem[1986]{lee}
Lee, H. M. \& Ostriker, J. P. 1986, ApJ 310, 176
\bibitem[1968]{lw}
Lynden-Bell, D. \& Wood, R. 1968, MNRAS 138, 495
\bibitem[1994]{maoz}
Maoz, E. 1994, ApJ 428, L5
\bibitem[1996]{ms}
Metcalf, R. B. \& Silk, J. 1996, ApJ 464, 218
\bibitem[1979]{miller}
Miller, G.E. \& Scalo, J.M. 1979, ApJS 41, 513
\bibitem[1995]{moore}
Moore, B. \& Silk, J. 1995, ApJ 442, L5 
\bibitem[1983]{Morrison}
Morrison, R. \&  McCammon, D. 1983, ApJ 270, 119
\bibitem[1989]{Murray} Murray, S. D. \& Lin, D. N. C. 1989, ApJ 339, 933
\bibitem[1995]{nakajima}
Nakajima, T. et al. 1995, Nat 378, 463
\bibitem[1995]{neuf}
Neufeld, D.A., Lepp, S. \& Melnick, G.J. 1995, ApJS 100, 132
\bibitem[1997]{nf}
Nulsen, P.E.J. \& Fabian, A.C. 1997, submitted to MNRAS
\bibitem[1994]{ODea}
O'Dea, C. P. et al. 1994, ApJ 422, 467 
\bibitem[1985]{osl}
Ostriker, J. P., Statler, T. S. \& Lee, H. M.  1985,
Bull.A.A.S. 16, 946
\bibitem[1983]{Palla}
Palla, F., Salpeter, E. E. \& Stahler, S. W. 1983, ApJ 271, 632
\bibitem[1988]{Palla1}  Palla, F \&  Stahler, E. E. 1988, 
in {\it Galactic and Extragalactic Star Formation}, ed. by 
R.E. Pudeitz and M. Fich, NATO ASI Ser. C {\bf 232}
(Kluwer Academic Publishers, Dordrecht, 1988)
p. 519. 
\bibitem[1990]{ps}
Persic, M. \& Salucci, P. 1990, MNRAS 247, 349
\bibitem[1994]{pfenniger}
Pfenniger, D., Combes, F. \& Martinet, L. 1994, A\&A 285, 79
\bibitem[1977]{press}
Press, W.H. \& Teukolsky, S.A. 1977, ApJ 213, 183
\bibitem[1996]{quinlan}
Quinlan, G. D. 1996, New Astronomy 1, 35
\bibitem[1995]{rebolo} 
Rebolo, R., Zapatero Osorio,  M. R. \&  Martin, E. L. 1995,
Nat 377, 129 
\bibitem[1997]{renault}
Renault, C. et al. 1997, A\&A 324, L69
\bibitem[1996]{rmv}
R\"ottgering, H., Miley, G. \& Van Ojik, R. 1996, The Eso Messenger No. 83, 26
\bibitem[1990]{ryu}
Ryu, M., Olive, K. \& Silk, J. 1990, ApJ 353, 81
\bibitem[1997]{sackett}
Sackett, P. D. 1997, ApJ 483, 103
\bibitem[1996]{saumon}
Saumon, D. et al. 1996, ApJ 460, 993 
\bibitem[1985]{Scalo} 
Scalo, J. M. 1985, in {\it Protostars and planets II},
ed. by D. C. Black and M. S. Mathews 
(University of Arizona Press, Tucson, 1985), p. 201
\bibitem[1969]{spitzer1969}
Spitzer, L. 1969, ApJ 158, L139
\bibitem[1987]{spitzer1}
Spitzer, L. 1987, {\it Dynamical Evolution of Globular Clusters},
Princeton University Press, Princeton
\bibitem[1971]{sh}
Spitzer, L \& Hart, M. H. 1971, ApJ 164, 399
\bibitem[1980]{Mathieu}
Spitzer, L \& Mathieu, R. D. 1980, ApJ 241, 618
\bibitem[1972]{st}
Spitzer, L. \& Thuan, T. X. 1972, ApJ 175, 31
\bibitem[1987]{soc}
Statler, T. S., Ostriker, J. P \& Cohn, H. N. 1987, ApJ 316, 626 
\bibitem[1985]{stodol}
Stodolkiewicz, J. S.  1985, IAU Symposium 113, {\it Dynamics of Star
Clusters}, ed. Goodman, J. \& Hut, P. (Reidel, Dordrecht)
\bibitem[1990]{tama}
Tamanaha, C. M., Silk, J., Wood, M. A. \& Winget, D. E. 1990,
ApJ 358, 164
\bibitem[1996]{uwg}
Unavane, M., Wyse, R. \& Gilmore, G. 1996, MNRAS 278, 727
\bibitem[1986]{vanalbada}
van Albada, T. S. \& Sancisi, R. 1986, {\it Phil. Trans. R. Soc.
London} A 320, 447
\bibitem[1986]{vand}
van Dishoeck, E. F. \& Black, J. H. 1986, ApJS 62, 109
\bibitem[1995]{Vietri} 
Vietri, M. \& Pesce, E. 1995, ApJ 442, 618
\bibitem[1977]{zahn}
Zahn, J. P. 1977, A\&A 57, 383
\end{thebibliography}
\end{document}